\documentclass[aps,12pt,reprint,amsmath,amssymb,amsfonts,aps,prb,twocolumn,floatfix,nofootinbib,showpacs,longbibliography,includegraphics]{revtex4-2}

\usepackage{mathtools}
\usepackage{braket}
\usepackage{bm}
\usepackage{booktabs}
\usepackage{multirow}
\usepackage[indent=15pt]{parskip}
\usepackage{graphicx} 
\usepackage{libertinus}  
\usepackage{libertinust1math}
\usepackage[T1]{fontenc}
\usepackage[colorlinks, allcolors=blue]{hyperref}
\usepackage[dvipsnames]{xcolor}
\usepackage{comment}

\begin{abstract}
Given that any subsystem of a closed out-of-equilibrium quantum system is an open quantum system, its dynamics (reduced from the full system's unitary evolution) can be either Markovian (memory-less) or non-Markovian, with the latter necessarily impeding the process of relaxation and thermalization. Seemingly independently, such non-ergodic dynamics occurs when an initial state has spectral weight on the so-called quantum many-body scar states, which are non-thermalizing eigenstates embedded deep in the spectrum of otherwise thermal eigenstates. In this article, we present numerical evidence that, in the class of systems which exhibit scars-induced entanglement oscillations, the presence of quantum scars is a microscopic ingredient that enables and enhances non-Markovianity of the dynamics of subsystems. We exemplify this with the PXP model and its deformations which either enhance or erase the signatures of scarred dynamics when quenched from simple product states with significant overlaps with the scarred states. The effect of thermalizing or scarring initial states is also similarly investigated. By probing information backflows with the dynamical behaviour of the distances between temporally-separated transient states of small subsystems, systematic signatures of subsystem non-Markovianity in these models are presented. It is seen that scarring-enhancing (erasing) deformations also exhibit enhanced (diminished) subsystem non-Markovianity. Likewise, results relating scarring (thermalizing) initial states to stronger (weaker) subsystem non-Markovianity are also presented. The retention of memory and revivals between transient subsystem states is a finer form of memory effect than captured by the revivals of full system's fidelity with the initial states. This sheds new light on the dynamical memories associated with quantum scarring, and opens interesting new questions at the interface of quantum scarring and an open quantum systems approach to investigating far-from-equilibrium and non-thermalizing isolated quantum many-body systems.
\end{abstract}

\begin{document}

\title{Quantum scarring enhances non-Markovianity of subsystem dynamics}

\author{Aditya Banerjee}
\email{adityabphyiitk@gmail.com}
\affiliation{Theory Division, Saha Institute of Nuclear Physics, 1/AF Bidhannagar, Kolkata 700064, India}
\affiliation{Centre for Quantum Engineering, Research and Education, TCG CREST, Sector V, Bidhannagar, Kolkata 700091, India}

\maketitle

\section{Introduction}     \label{sec1}

Non-integrable quantum many-body systems (QMBS) that are far away from equilibrium are expected to thermalize when their Hamiltonian evolutions are initiated from generic, \textit{typical} initial states \cite{Gogolin2015}. By "typical", one means an initial state that has a well spread support over the eigenstates of the Hamiltonian in an appropriate microcanonical window and ideally has approximately equal support on a majority of those eigenstates. For a QMBS that can thermalize in principle, its eigenstates which are sufficiently higher in the spectrum from the ground state(s) are volume-law entangled and satisfy the Eigenstate Thermalization Hypothesis (ETH) \cite{Deutsch2018}. Such eigenstates are often referred to as thermal eigenstates. 

Notwithstanding this typical picture, in recent years an atypical situation of special interest is that of the so-called quantum many-body scar states which lead to a weak-breaking of ergodicity \cite{Serbyn2021,Chandran2023}. These are high energy eigenstates in the bulk of the spectrum of a QMBS that are \textit{non-thermal} : they are usually entangled sub-extensively, violate the ETH and when an initial state has a notable overlap with them, non-thermalizing dynamics follows. These scar states and their central role in non-thermalizing dynamics, by way of notable overlaps with simple and experimentally feasible initial states, thus manifests a concrete scenario that breaks certain crucial conditions required for equilibration and thermalization from basic principles as established in Refs.\cite{Reimann2008,Linden2009,Short2010,Short2011}. What is remarkable is that these non-thermal scar states exist amongst thermal states in the spectrum, and generally speaking the number of these states is only polynomial in system size and therefore exponentially suppressed relative to the remainder of the spectrum. Furthermore, the quality of fidelity revivals is naturally dependent on how significantly overlapping a given initial state is with the set of scars in the system's spectrum. Thus, when these overlaps are not particularly significant yet are not negligible, one expects thermalization to eventually follow within experimental or computational times after a prolonged "pre-thermalization" regime. Indeed, for systems in the thermodynamic limit, perfect fidelity revivals are an exceptionally rare occurrence in general, and any scarred dynamics can only result in a slower approach to an otherwise guaranteed thermalization in the long-time limit. The actual attainment of thermal equilibrium for sufficiently large systems in some cases can be well beyond computational and experimental times and resources, however. 

It happens with many scarred models that their scar states are approximately equally spaced in energy and so the scarred subspaces take the form of towers \cite{Moudgalya2022,Chandran2023}. When an initial state in a quench dynamics of these systems happens to be a near-perfect superposition of these tower states, the resultant dynamics is essentially restricted to the scarred subspace, and thus becomes practically disconnected from the rest of the Hilbert space. In that case, one expects entanglement entropies (and also physical observables) to exhibit long-lasting oscillations. However, it turns out that this does not happen for most known models of scarred QMBS, owing to a recently proposed no-go theorem that seems to govern most of these cases \cite{Odea2025}. As such, scarred QMBS with oscillatory entanglement dynamics are a minority and therefore of special interest. The most well-known quantum scarred model, the PXP model, exhibits imperfect signatures of scarred dynamics in its base form, but these become stronger and more persistent with the inclusion of certain quasi-local terms in the PXP Hamiltonian with tunable and optimal coefficients \cite{Choi2019}.

A different aspect of the non-equilibrium dynamics of a closed QMBS is the (non-)Markovianity of the dynamics of its subsystems, which however can also depend on the size of the subsystems in question in addition to other physical parameters of a given system \cite{Banerjee2025a}. This is rooted in two essential facts. Firstly, any subsystem of a closed out-of-equilibrium QMBS is a bonafide open quantum system interacting with an environment which is the remainder part of the full system. And secondly, the dynamics of any open system interacting with an environment can be characterized as either memory-less (Markovian) with a monotonic loss of information out of the system into its environment, or it can retain memory of its past due to information backflows from the environment back to the system (non-Markovian) \cite{Vacchini2024}. Let us describe the scenario briefly. When the coupling between a quantum system in question and its environment is weak (in comparison to the interaction timescales of the environment, in which case the environment can act as a "bath" in self-equilibrium unaffected by the dynamics of the quantum system), the dynamics is generically Markovian. This is because the bath, owing to it being unaffected by its coupling to the quantum system in question, instead drives the quantum system to an equilibrium state compatible with its own equilibrium (the bath provides the temperature to which the quantum system should equilibrate to - this is thus ultimately going to be described by the canonical ensemble picture of statistical mechanics). In such a dynamics, the quantum system \textit{monotonically loses memory} of its past states, with a one-way flow of such information from the system out into the bath, but since the bath is unaffected by the dynamics of the system at hand, such information flows are lost forever (information is not a conserved quantity). As such, the future state of the quantum system is only dependent on its present state and not on the past states. However, when the system and its environment are strongly coupled such that the environment is affected by the system's dynamics and its intrinsic correlation timescales are comparable to their coupling strength, the system's dynamics is typically non-Markovian (or equivalently, one says that the environment is non-Markovian). In this case, the information loss is not monotonic, meaning that there is usually an information backflow from the environment back into the system, and thus the dynamics of the quantum system in question depends on its past states in a complicated manner and the system develops temporal correlations between past-future states \cite{Rivas2014,Breuer2016,Vega2017,Li2018}. In the strongly non-Markovian cases, both the system and its environment behave essentially on the same footing and both remain far out of equilibrium for long times.

Given then that any subsystem of a given closed QMBS is an open quantum system, it is pertinent to ask if its reduced (from the full system's unitary Hamiltonian evolution) dynamics is Markovian or non-Markovian. This can be investigated using several quantum informational notions commonly used in open quantum systems theory to ascertain the (non-)Markovianity of a given dynamical system. We believe that characterizing the subsystem dynamics of closed QMBS as Markovian or non-Markovian provides an additional insight into the non-equilibrium dynamics of closed QMBS. In Ref.\cite{Banerjee2025a}, using trace distance based measures of quantum distances between \textit{temporally-separated} states of small subsystems, we characterized the (non-)Markovianity of subsystem dynamics induced by strong quenches far across the critical point in the paradigmatic mixed field Ising spin chain. We remark here that subsequent to the publication of Ref.\cite{Banerjee2025a}, we became aware of an earlier related work of Ref.\cite{Rajagopal2010} which also considered detecting and quantifying non-Markovianity in terms of non-contractivity of distinguishability (as measured by the fidelity) between temporally-separated transient states of an open quantum system.

In this work, within the scenario where entanglement entropies are not frozen and exhibit oscillations, we are interested in investigating the relationship between quantum scarring and non-Markovianity of (small) subsystems. Specifically, we carry out our work with the PXP model and some of its deformed variants that are known to enhance or diminish central qualities of scarring dynamics, as well as with varying initial states that are known to have notable (or lack thereof) overlap with PXP model's scars thus leading to scarring (or thermalizing) dynamics. It will be seen that the deformations of the PXP model that stabilize and enhance scarring dynamics also exhibit enhanced levels of non-Markovianity of subsystem dynamics, and the opposite happens with the deformations that work to erase scarring dynamics and restore ergodicity. Likewise, it will be seen that the scarring or thermalizing initial states respectively lead to pronounced subsystem non-Markovianity or lack thereof. These features are of course expected for a system that exhibits global revivals and long-lived-oscillatory behaviour of observables, and thus establishes the internal consistency of looking at such dynamics through the lens of subsystem non-Markovianity. Indeed, the highly constrained global system dynamics prohibits local subsystems to sufficiently entangle with their complement and consequently fail to lose their memories, leading to strongly non-Markovian dynamics.

The focus on small subsystems comprising of a few spins is partly driven by the numerical costs of constructing sufficiently large density matrices, partly with an intention to obtain a picture of the dynamics at the "fine-grained" level of a few individual spins, and partly due to remarkable site-resolved experimental control in practice in ultracold experimental platforms (see e.g. \cite{Ott2016,Kuhr2016,Gross2021}) which render theoretical questions and results regarding the dynamics of a few spins in a large system experimentally meaningful. We point out here that a recent work \cite{Perciavalle2025} has studied a potentially related notion of "local reminiscence" of small subsystems in the PXP model, i.e., retention of memory of the initial state in the transient states of the subsystems in question.

This article is organized as follows. In Section.\ref{sec2}, we set our notations and conventions, and provide an introduction to the notion of subsystem non-Markovianity and information backflows. Section.\ref{sec3} introduces the PXP model and its PXPZ and PXPXP deformations that are studied in this work, and the results and interpretations are presented in Section.\ref{sec4}. The main body of this work concludes in Section.\ref{sec5}. The appendices present example demonstrations of numerical benchmarks, and additional results related to the dynamics of entanglement negativity, an observation on differently configured subsystems, and classical signatures of subsystem non-Markovianity. 

\section{Overview of requisite notions}  \label{sec2}

Quantum states are represented by their density matrix operators, which are Hermitian matrices with positive eigenvalues and unit trace. Given the set of density matrices $ \mathcal{D}(\mathcal{H})$, a trace-preserving quantum operation $\Lambda$ is termed positive if $\Lambda \! : \! \mathcal{D}(\mathcal{H}) \! \rightarrow \! \mathcal{D}(\mathcal{H})$. Moreover and more importantly, when the operation $\mathbb{I} \otimes \Lambda$ is positive (here the support of the identity operation $\mathbb{I}$ is complementary or ancillary to the support of the action of $\Lambda$), the quantum operation $\Lambda$ is said to be a completely-positive (CP) and trace-preserving (CPTP) map (also commonly known as quantum channels). Some concrete examples include unitaries, measurements and partial trace. Any physically realizable quantum operation necessarily has to be CPTP, as states/density matrices can only be converted to other states/density matrices in the laboratory by local quantum operations.

\textit{Distances between quantum states and CPTP contractivity---} It is possible to quantify how "far" (or distinguishable) quantum states are from each other in the Hilbert space. Many measures exist in the literature of quantum information theory for this purpose, each with different operational interpretations and advantages over the others (see e.g.\cite{Wilde2017,Hayashi2017}). We choose the trace distance as the measure of distance between quantum states in this work (results can be verified to be qualitatively unchanged if other measures are chosen). We make this choice due to the simplicity of computability of the trace distance, because it does not involve any fractional powers or logarithms of  density matrices (which can give rise to numerical instabilities, non-uniqueness and related distracting issues for density matrices that are numerically non-invertible or close to being so). Given any two density matrices $\rho$ and $\sigma$ (assumed to be of same rank), the trace distance (TD) between them is defined as,
\begin{equation}     \label{TD}
    T_d(\rho,\sigma) = \frac{1}{2}\sum|\lambda_i|  \text{    ,}
\end{equation}
where $\{\lambda_i\}$ is the set of eigenvalues of $(\rho-\sigma)$, with the index $i\!=\!1,2,..,\operatorname{rank}(\rho-\sigma)$. It takes values $\in [0,1]$, with the value $0$ signifying the two states in question operationally indistinguishable from each other and the value $1$ likewise signifying maximally-distant/distinguishable pair of states in the Hilbert space. While other metrics (normalized to lie in $\in [0,1]$ if not already so by definition) of distances between quantum states may yield different values of the distance between any given pair of states, they must agree on when they are fully indistinguishable (the value $0$) or maximally distinguishable (the value $1$).

A fundamental property of the trace distance (or any other distance measures) is that it is non-increasing or \textit{contractive} under the action of CPTP maps \cite{Wilde2017,Hayashi2017}: given a CPTP operation $\Lambda:\mathcal{D(\mathcal{H})} \! \rightarrow  \! \mathcal{D(\mathcal{H})}$, the trace distance satisfies,
\begin{equation}      \label{TDCPTP}
    T_d(\Lambda(\rho),\Lambda(\sigma)) \leq T_d(\rho,\sigma)    \text{   .}
\end{equation}
The message carried by this inequality is that physically realizable quantum operations can not increase the (obtainable information about the) distance/distinguishability between any given pair of quantum states.

\textit{Information backflow and non-Markovianity---} Let $\rho_t^{(\ell)}$ be the (reduced) density matrix at time $t$ of a subsystem of length $\ell$ (i.e., it contains $\ell$ spins), with $t\!=\!0$ denoting the starting time of the Hamiltonian dynamics. We are interested in the question of how the (trace) distance between two transient states separated in time by an amount $\delta$ evolves with time, i.e., we wish to investigate the behaviour of $T_d(\rho_{t+\delta}^{\ell}, \rho_t^{\ell})$ as a function of time $t$, for a given temporal separation $\delta$, as well as the change of this behaviour with varying $\delta$. Given that the subsystem in question is an open quantum system, one can formally obtain its "reduced" dynamics from the full system's unitary dynamics. However, actually deriving (and then solving) the exact reduced dynamical evolution equations for subsystems is a very non-trivial and complicated task. However, the formalism of quantum dynamical semigroups \cite{Alicki2007,Chruscinski2022} allows to classify some dynamics as Markovian or non-Markovian based on simple formal considerations without necessarily computing the reduced dynamical evolutions. Given the subsystem's state at time $t_0$ and when the subsystem and its environment are initially uncorrelated (such as product states), let us denote by $\{\Lambda_{t-t_0}\}$ the family of dynamical CPTP maps that evolve the subsystem's state at time $t_0$ to time $t$, i.e., 
\begin{equation}    \label{map}
    \rho_{t}^{\ell} = \Lambda_{t-t_0}[\rho_{t_0}^{\ell}]  \text{   .}
\end{equation}
In particular, when the Hamiltonian is time-independent, one can always set $t_0\!=\!0$, leading to $\rho_{t}^{\ell} = \Lambda_{t}[\rho_{0}^{\ell}]$. The semigroup property $\Lambda_{t+s} \!=\! \Lambda_{t}\Lambda_{s}$ (for any $t,s \geq 0)$ (this is also an instance of the CP-divisibility property, see e.g. \cite{Rivas2014,Breuer2016,Chruscinski2022}) can then be used to write $\Lambda_t \!=\! (\Lambda_1)^t  $, in terms of the most elementary dynamical propagator $\Lambda_1$ that propagates a given state by a unit time-step. Satisfaction of this semigroup property by a given family of quantum dynamical maps is often taken as a definition of Markovian dynamics or of the family of dynamical maps being Markovian (and its violation as a definition of non-Markovianity \cite{Rivas2014,Breuer2016}). This also implies that given the state $\rho_{\delta}^{\ell} \!=\! \Lambda_{\delta}[\rho_0^{\ell}] \!=\! (\Lambda_1)^{\delta}[\rho_0^{\ell}]$ at time $t\!=\!\delta$, its evolved state at time $t\!=\!t+\delta$ is $\rho_{t+\delta}^{\ell} \!=\! \Lambda_{t+\delta}[\rho_0^{\ell}] \!=\! \Lambda_t[\rho_{\delta}^{\ell}]$. From Eq.(\ref{TDCPTP}) then, a non-increasing behaviour of the trace distance $T_d(\rho_{t+\delta}^{\ell}, \rho_t^{\ell})$ follows,
\begin{equation}             \label{tdcptp1}
    T_d(\rho_{t+\delta}^{\ell}, \rho_t^{\ell}) = T_d(\Lambda_t[\rho_{\delta}^{\ell}],\Lambda_t[\rho_{0}^{\ell}]) \leq T_d(\rho_{\delta}^{\ell},\rho_{0}^{\ell})      \text{   .}
\end{equation}
When these inequalities are violated at any time $t$ or at any temporal separation $\delta$, it is a signature of the violation of the semigroup property of Markovian evolutions, and therefore a defining signature of non-Markovianity. The violation of these inequalities is physically interpreted as backflows of information (about the distinguishability between quantum states) from the environment back to the subsystem in question, and this defines a separate notion of non-Markovianity \cite{Breuer2009} (see however a recent work relevant to this notion \cite{Buscemi2025}). Generally speaking, information backflows into the subsystem occur when its environment is not able to function as a monotonic absorber or a "bath" of the information about distinguishabilities between the states of the subsystem in question. This typically happens beyond the standard weak-coupling picture of open quantum systems such as the Born-Markov framework \cite{Vacchini2024}. In particular, when the effective coupling between a subsystem and its environment is stronger than or comparable to the intra-environment interaction scales, information backflows are expected to occur generally. 

As advocated for in Ref.\cite{Banerjee2025a}, this information backflow notion of non-Markovianity is the one we work with directly, as this simply requires comparison of the trace distances as per Eq.(\ref{tdcptp1}) without having to extract \textit{all} the dynamical maps and verifying their semigroup property or lack thereof, which is a separate problem in itself. However, the appearance of information backflows (violations of the above inequalities) is evidence of the violation of the semigroup property of the dynamical maps (whatever their actual mathematical form may be). More importantly, contractivity of the trace distances (Eqs.(\ref{TDCPTP},\ref{tdcptp1})) holds for the larger and more general case of positive maps (of which the CP maps are a subset) \cite{Ruskai1994}, and thus non-Markovianity as defined by the non-contractivity of trace distances is a stronger criteria \cite{Chruscinski2011,Chruscinski2022}.

One can define a degree of non-Markovianity by the total magnitude of such violations. In a discrete time dynamics, as in our numerical simulations with a fixed time-step $\tau$, considering the discrete "slope" of the trace distance for a given $\delta$ (note that this is always negative for Markovian dynamics), 
\begin{equation}    \label{slope}
    \alpha(t,\delta) = \frac{1}{\tau}\bigg(T_d(\rho_{t+\tau+\delta}^{\ell}, \rho_{t+\tau}^{\ell}) - T_d(\rho_{t+\delta}^{\ell}, \rho_{t}^{\ell}) \bigg)   \text{    ,}
\end{equation}
a degree of non-Markovianity can be constructed with the cumulative magnitude of revivals (increases) of the trace distance $T_d(\rho_{t+\delta}^{\ell}, \rho_t^{\ell})$,
\begin{equation}     \label{TDdegree}
    \mathcal{D}(\delta) = \sum_t \alpha(t,\delta)  \hspace{0.3cm} \forall t \hspace{0.25cm} \text{s.t.} \hspace{0.25cm} \alpha(t,\delta)>0    \text{ .}
\end{equation}
Note that this is not the same as the degree of non-Markovianity defined in \cite{Breuer2009,Breuer2010}, where a maximization over pairs of initial states was made as they wanted to define a non-Markovianity degree for a given family of quantum dynamical maps acting on a set of different initial states. But our concern is slightly different, because we wish to work with a fixed initial state for a given class of quenching, and then to quantify the levels of non-Markovianity resulting from quenching the given initial state by different Hamiltonians in the PXP family.

\section{The PXP model and its deformations}    \label{sec3}

The kinetically constrained PXP model achieved prominence ever since it was shown to describe the experimentally observed constrained dynamics of Rydberg atom based quantum simulators \cite{Bernien2017}. The constraint it imposes is of dynamically freezing any two neighboring spin ups $\mid\uparrow\uparrow\rangle$. It hosts non-thermalizing scar states in its spectrum \cite{Turner2017,Turner2018}, which leads to revivalist quench dynamics when initiated from initial states with reasonable overlap with the scarred states, the most simple and implementable among such initial states is the Néel product state.

\begin{figure}[!htbp]
	\centering
	\includegraphics[width=4.25cm,height=4.2cm]{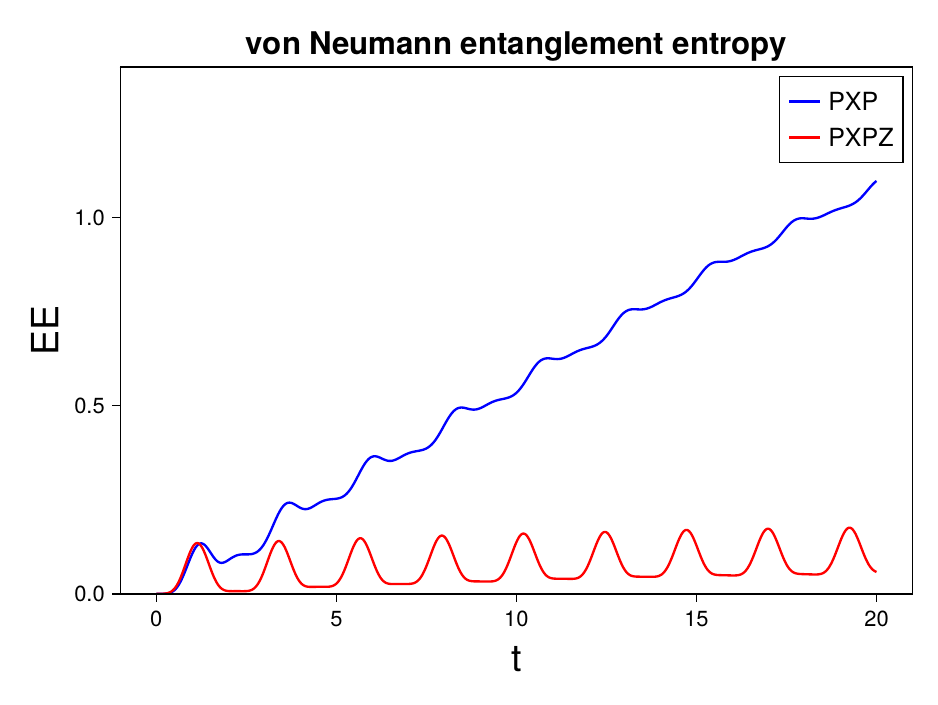}
        \includegraphics[width=4.25cm,height=4.2cm]{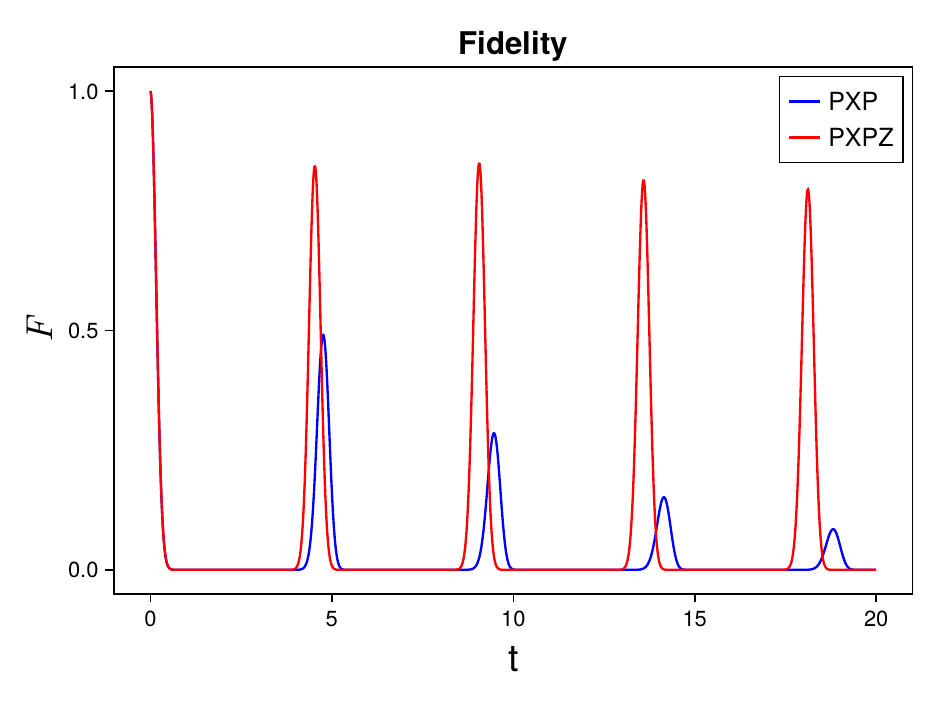}
        \includegraphics[width=4.25cm,height=4.2cm]{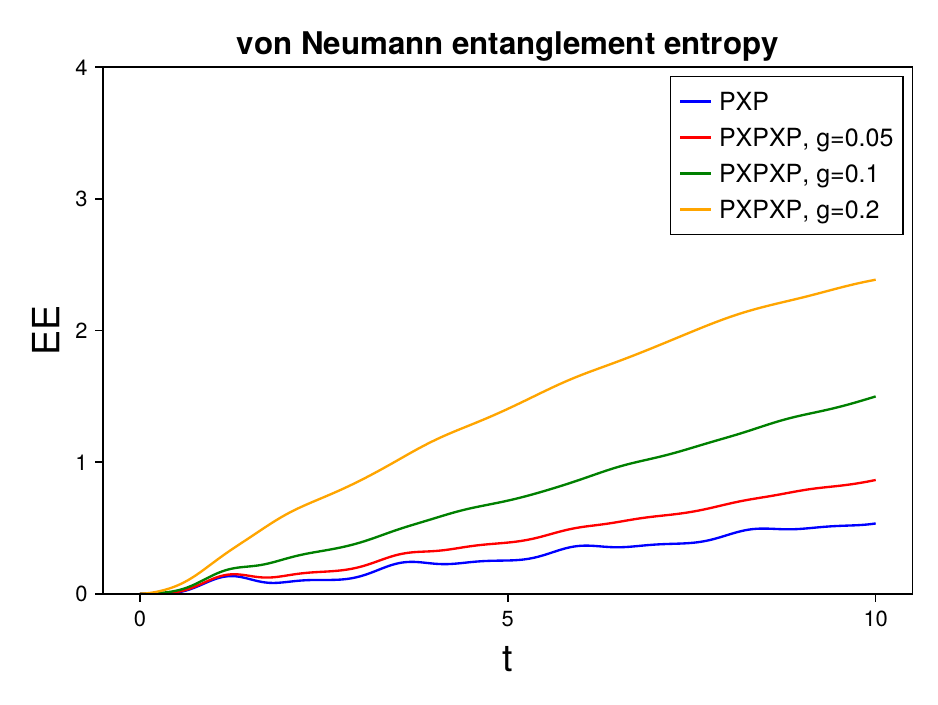}
        \includegraphics[width=4.25cm,height=4.2cm]{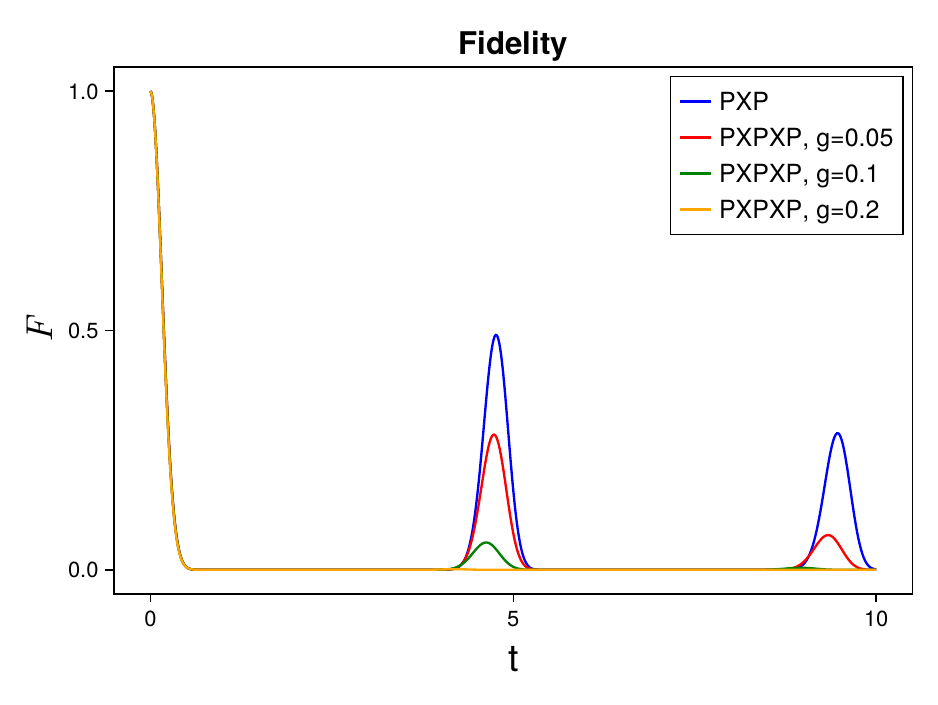}
        \includegraphics[width=4.25cm,height=4.2cm]{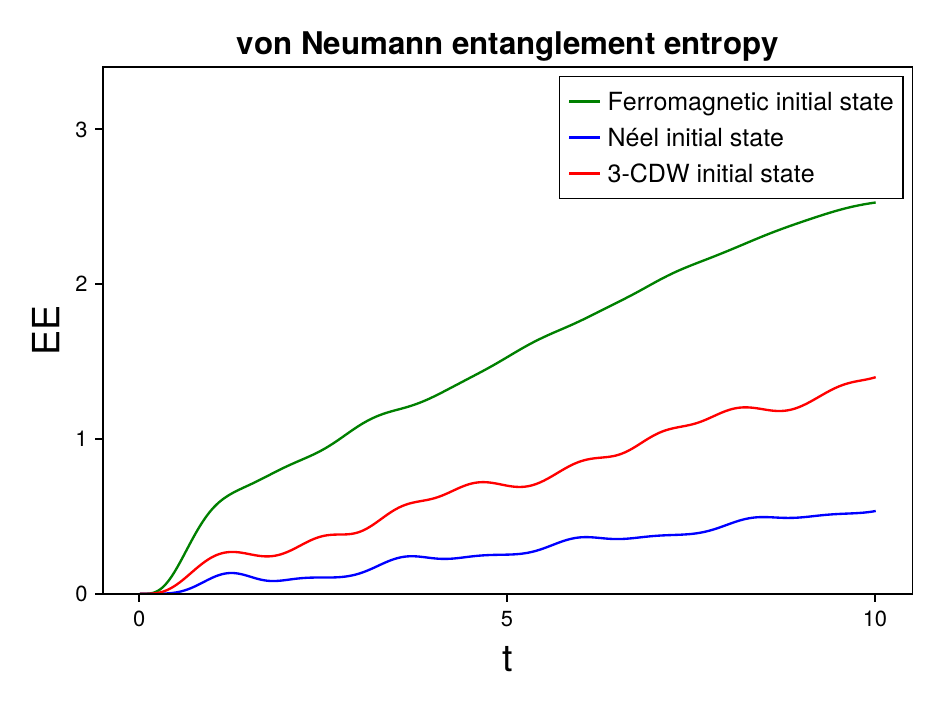}
        \includegraphics[width=4.25cm,height=4.2cm]{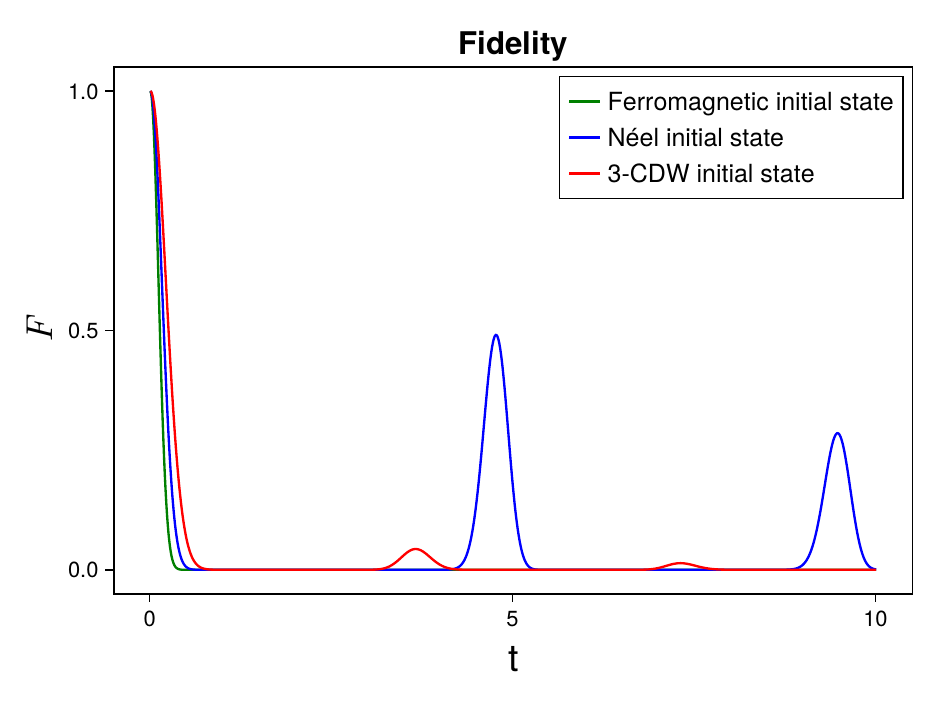}
	\caption{\fontsize{9}{11} \selectfont A review of known results shown here for completeness. \textbf{(Upper row, left)} Half-chain von Neumann entanglement entropy $EE\!=\!-\operatorname{tr}\rho\operatorname{ln}(\rho)$ where $\rho$ is the reduced density matrix of the halved subsystem, and \textbf{(Upper row, right)} fidelity $F\!=\!|\langle\Psi_t | \Psi_0 \rangle |^2$ for the PXP vs. PXPZ models, where $\Psi_0$ is the Néel initial state and $\Psi_t$ is the full system transient (pure) state. The fidelity revivals in the PXP model appear at time intervals $\approx 4.76 $ while in the PXPZ model they appear at intervals $\approx 4.52$. PXPZ deformation leads to strongly oscillatory entanglement entropy, and stronger and more persistent fidelity revivals, signifying enhancement of PXP scars. \textbf{(Middle row)} The same quantities for the PXPXP deformation shown at increasing strengths $g$, compared with the base PXP model. This deformation shows increasingly opposite behaviour compared to that of PXPZ, signifying signatures of erasing the scars away, thus restoring ergodicity and quicker thermalization. \textbf{(Lower row)} Comparison of the same quantities in the case of PXP quench from ferromagnetic, Néel ($\mathbb{Z}_2$) and 3-CDW ($\mathbb{Z}_3$) initial states.}
	\label{fig:fig1}
\end{figure}

The base PXP model is defined by the Hamiltonian (in open boundary conditions),
\begin{equation}    \label{PXP}
    \mathcal{H}_{PXP} = \sum_{i=2}^{N-1} \mathcal{P}_{i-1}\sigma_i^x\mathcal{P}_{i+1}  +   \text{   boundary terms              ,}
\end{equation}
where $\mathcal{P}_i \!=\!(\mathbb{I} - \sigma_i^z)/2$ projects onto spin-down $\mid\downarrow\rangle$ subspace at each site $i$. In the above, we have suppressed writing the boundary terms $(\sigma_1^x\mathcal{P}_2+\mathcal{P}_{N-1}\sigma_N^x)$ but these are included in our simulations. Due to this projection, a spin-up or -down at a site $i$ is flipped only if both of its neighbors are in spin-down configuration. This is the kinetic constraint of this model due to which the effective/reduced Hilbert (sub)space is much smaller than the full Hilbert space and results in dynamics occurring within this reduced subspace of the full Hilbert space. It was shown in Refs.\cite{Turner2017,Turner2018} that this model hosts low-entangled scarred states in its spectrum, exhibits linear growth (with a rate which is slower than in efficiently-thermalizing systems, and dressed with mild oscillations) of von Neumann entanglement entropy as well as partial fidelity revivals when initiated from the simple $\mathbb{Z}_2$ (Néel) product state $\mid\uparrow\downarrow\uparrow\downarrow...\rangle$ and its inverted counterpart, which were shown also to have more overlap with the set of scarred eigenstates compared to the other states in the spectrum. In particular, Refs.\cite{Turner2017,Turner2018} also showed how to analytically construct these scar states exactly or approximately using the so-called forward scattering approximation. Its non-thermalizing dynamics was also argued to be proximal to some nearby integrable model \cite{Khemani2019}. Subsequently, some exact area-law-entangled scarred states were found in this model \cite{Lin2019,Ivanov2025}, and surprisingly even those with volume-law entanglement \cite{Ivanov2024, Mohapatra2025}. Approaches to understand quantum scarring in constrained systems based on classical phase space pictures in partial analogy to single-particle scars were studied in \cite{Ho2019,Michailidis2020}, and more recently an approach to understand the dynamical revivals typical of scarred systems based on a broken-unitary picture was proposed in \cite{Rozon2024} for PXP and several other systems. 

Moreover, when initiated from the fully polarized ferromagnetic state $|\downarrow\downarrow...\rangle$, the PXP quench dynamics is featureless and thermalizing. Whereas when the PXP quench is initiated from the $\mathbb{Z}_3$ state $|\uparrow\downarrow\downarrow\uparrow\downarrow\downarrow...\rangle$ (which we refer to as 3-CDW (three-site charge density wave) state), one observes noticeably weakened scarred dynamics compared to the Néel case. The contrast in the behaviour of full system fidelity and half-chain von Neumann entanglement entropy, with respect to the three initial states, for the PXP quench is displayed in Fig.\ref{fig:fig1} (lower row).

In Ref.\cite{Choi2019}, it was numerically shown that the fidelity revivals, (when initiated from the Néel state) and the scarred states of the PXP model in Eq.(\ref{PXP}) could be made more prominent and perfect by perturbatively deforming it with certain longer-ranged terms in the Hamiltonian of the form,
\begin{equation}    \label{PXPZ}
    \Delta\mathcal{H}_r = -\lambda\sum_i  \mathcal{P}_{i-1}\sigma_i^x\mathcal{P}_{i+1} (\sigma_{i-r}^z+\sigma_{i+r}^z)        + \text{ boundary terms    ,}
\end{equation}
where $r\!=\!\{2,3,..\}$. We will refer to the PXP model with this deformation as the PXPZ model, $\mathcal{H}_{PXPZ} \!=\! \mathcal{H}_{PXP} + \sum_r\Delta\mathcal{H}_r$. A forward scattering approximation calculation showed that the optimal value of the parameter $\lambda$ to be $\sim 0.05$. It is interesting that at about half this value and for $r\!=\!2$, the entire spectrum of the PXPZ model comes closest to an integrable-like least-thermal spectrum \cite{Khemani2019}. We focus, for simplicity, on the case $r\!=\!\{2,3\}$ in this work (that is, $r$ takes values $2$ and $3$), as the case of considering only $r\!=\!2$ distinguishes itself quite nominally from the base PXP model in our investigation of subsystem non-Markovianity, whereas including the case $r\!=\!3$ already shows considerably more distinguishing behaviour compared to the base PXP model. This also keeps the model sufficiently quasi-local compared to the most general case of unrestricted values of $r$ which leads to a very long-range Hamiltonian. Once again, the boundary terms have not been displayed in the above equation but have been included in our calculations. A comparison of fidelity revivals and half-chain entanglement entropy between the PXP and PXPZ models is shown in Fig.\ref{fig:fig1} (upper row). We note that this is not an original result and is being shown here for the sake of completeness. As seen in Fig.\ref{fig:fig1} (upper row), the von Neumann entanglement entropy grows slowly with a linear envelope dressed with oscillations, whereas that of the PXPZ model hardly grows at all and is strongly damped by persistent oscillations. The fidelity revivals seen in the right figure of Fig.\ref{fig:fig1} are equally striking, with the PXP model showing rather subdued revivals (at time periods $\approx 4.76$) which gradually but quickly diminish in their strengths, whereas that of the PXPZ model with $r\!=\!3$ shows almost persistent revivals to about $80-85\%$ strengths with a time period $\approx 4.52$, and this strength diminishes with time very weakly within our simulation times. Such contrasting behaviour is also seen in local density of states related to the decay rate of fidelity \cite{Schnee2025}. In Appendix-\ref{appn2}, expectedly more persistently oscillating behaviour of a mixed-state entanglement measure (associated with subsystems comprising of a few spins) is seen in the PXPZ case as compared to the PXP case. 

\begin{figure}[!htbp]
	\centering
	\includegraphics[width=5.6cm,height=5cm]{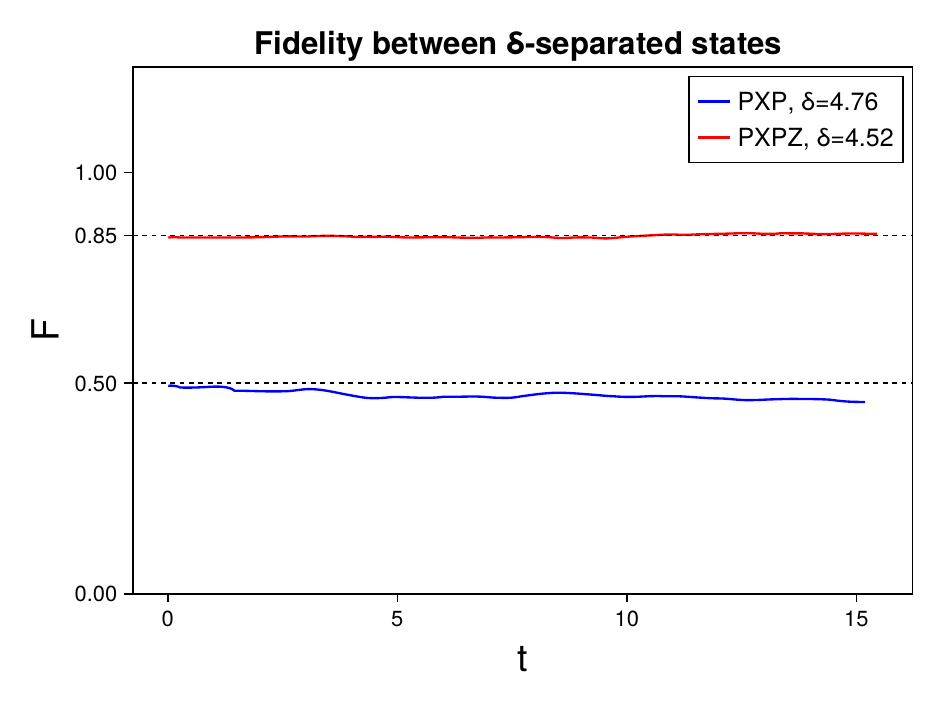}
	\caption{\fontsize{9}{11} \selectfont The fidelity $F\!=\!|\langle\Psi_{t+\delta} | \Psi_t \rangle |^2$} between transient full-system states separated in time from each other by $\delta\!=\!4.76$ for the PXP model, and $\delta\!=\!4.52$ for the PXPZ model. Approximately $85\%$ fidelity is retained in the PXPZ case, where as in the PXP model, it slowly decreases on average (but mildly non-monotonically) after an initial fidelity of $ \approx 49\%$.
	\label{fig:fig1a}
\end{figure}

Given the fidelity revivals of the transient full-system states with respect to the initial state at the aforementioned time periods, and the fact that the early-time dynamics is significantly dominated by the scarred subspace (even more so in PXPZ model), we expect non-zero fidelity between transient full-system states that are separated from each other in time by $\delta\!=\!4.76$ in the PXP case and $\delta\!=\!4.52$ in the PXPZ case. This is confirmed in Fig.\ref{fig:fig1a}, which further shows that this fidelity remains approximately constant in the PXPZ case while showing visibly, albeit very slowly, decreasing trend in the PXP case. This is another aspect of the long-time enhancement of high fidelity scarring dynamics induced by the PXPZ deformation, signifying more strongly subspace-restricted dynamics in this case compared to that in the base PXP model.

\begin{figure*}[!htbp]
	\centering
	\includegraphics[width=5.6cm,height=5cm]{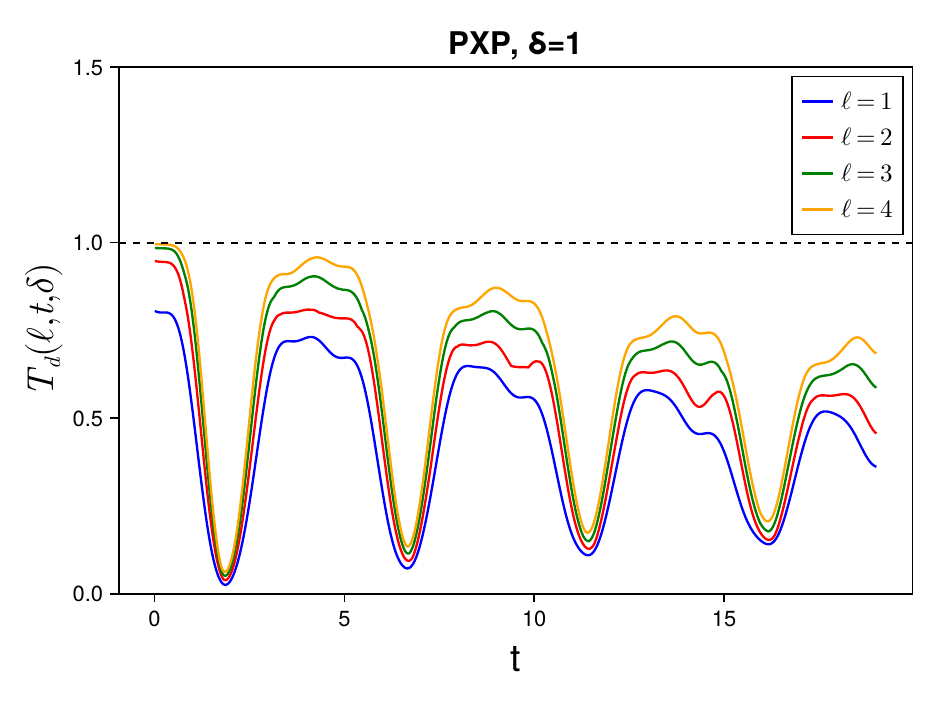}
        \includegraphics[width=5.6cm,height=5cm]{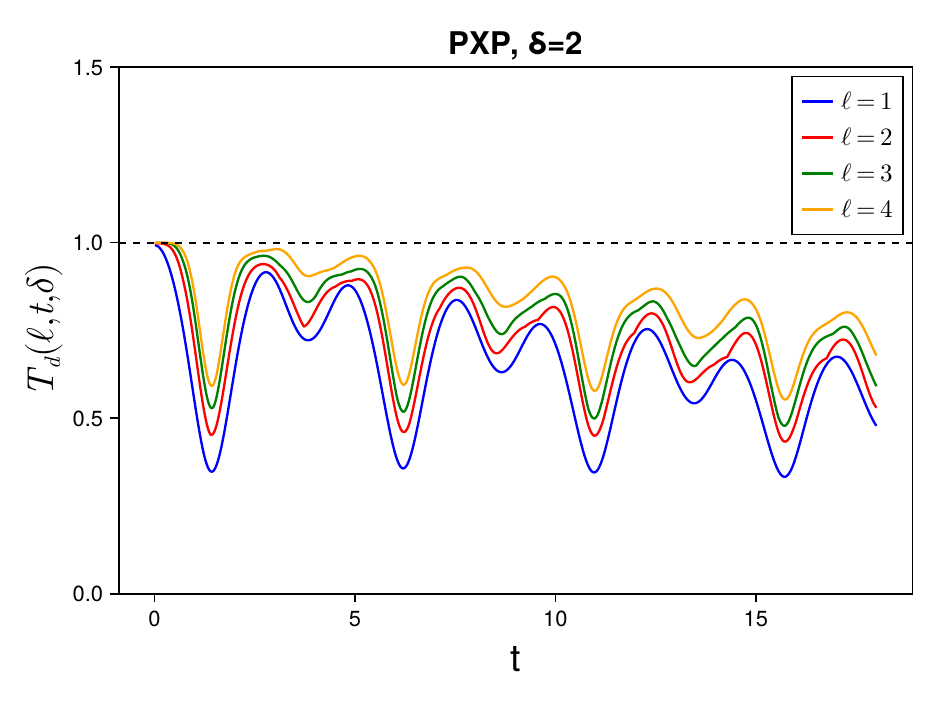}
        \includegraphics[width=5.6cm,height=5cm]{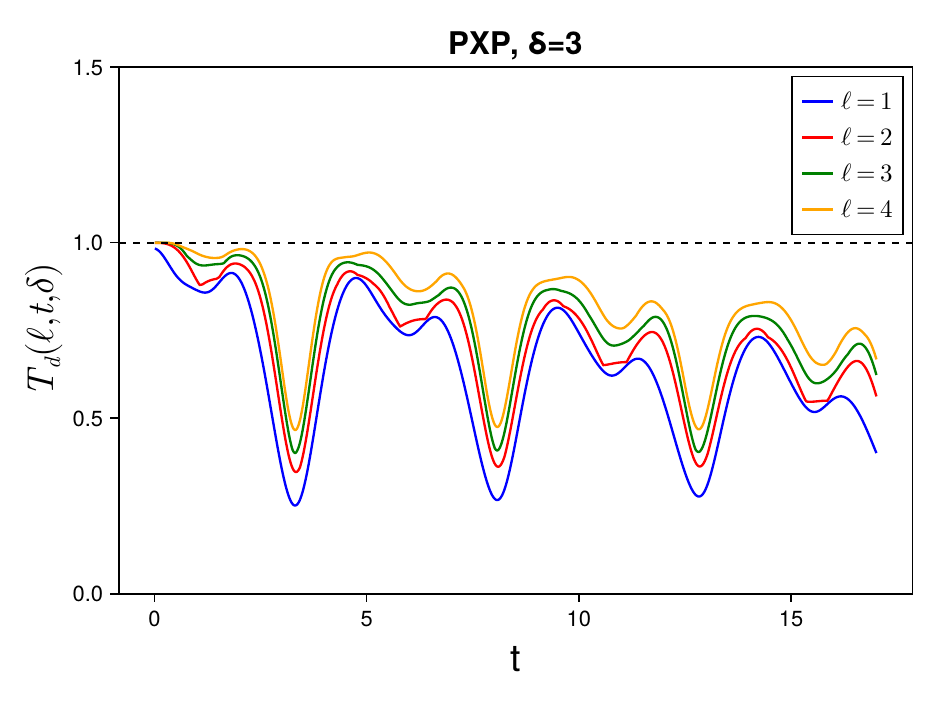}
        \includegraphics[width=5.6cm,height=5cm]{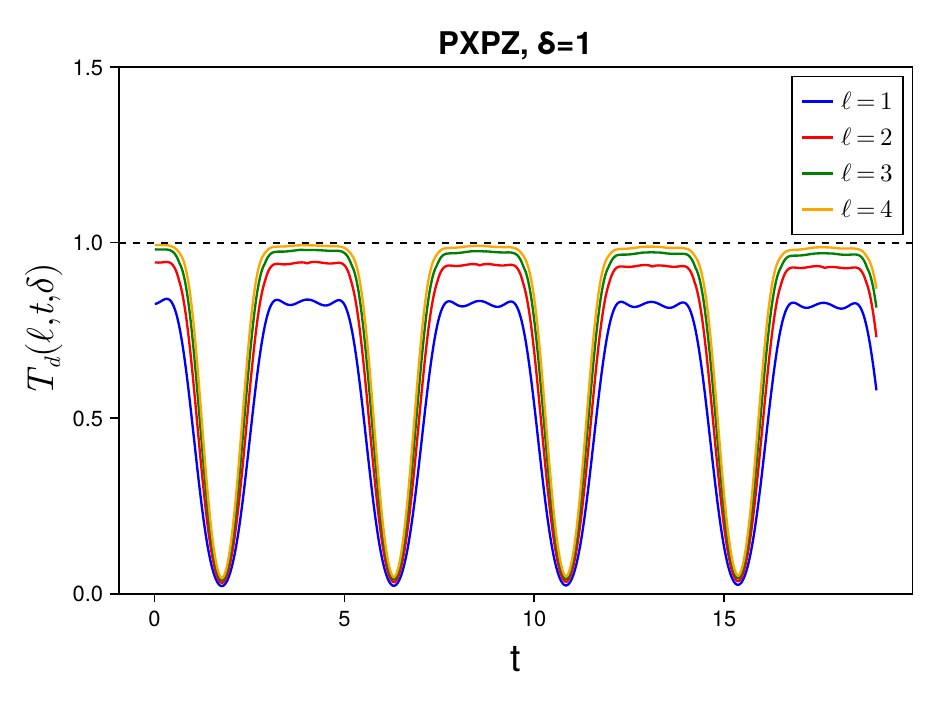}
        \includegraphics[width=5.6cm,height=5cm]{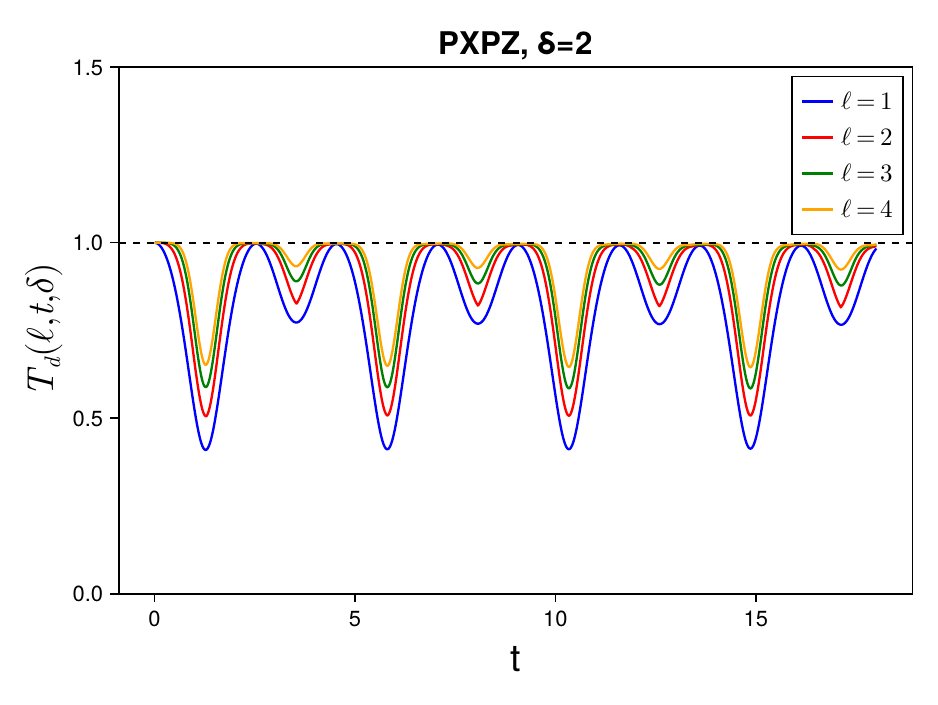}
        \includegraphics[width=5.6cm,height=5cm]{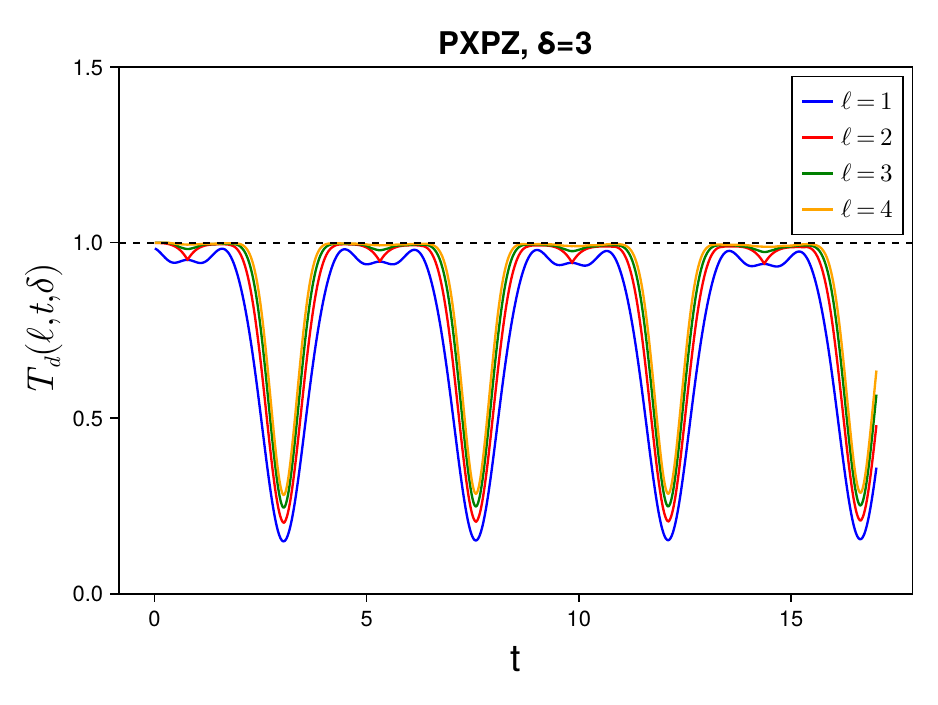}
	\caption{\fontsize{9}{11} \selectfont Behaviour of the trace distances $T_d(\ell,t,\delta)$ between $\delta$-separated subsystem states for various subsystems at three particular values of $\delta\!=\!\{1,2,3\}$, in the PXP model (\textbf{upper row}) and the PXPZ model (\textbf{lower row}). Persistently oscillatory behaviour is seen in the PXPZ model, whereas a slowly decaying profile is clear in the PXP model. Subsystem-relative strengths of the non-monotonic dynamics of $T_d(\ell,t,\delta)$ is dependent on $\delta$, such as when $\delta\!=\!1$, larger subsystems show larger non-monotonicities, whereas the opposite occurs when $\delta\!=\!2$. The deep minima occur at time periods $\approx 4.76$ in the PXP model and $\approx 4.52$ in the PXPZ model, however the actual instants in time $t$ is $\delta$-dependent, and the values attained in these minima is dependent both on $\delta$ and subsystem size $\ell$. As mentioned in the main text, for subsystems of size $\ell>1$, between any consecutive constituent spins in these subsystems there is a separation of one site.}
	\label{fig:fig2}
\end{figure*}

Importantly, Ref.\cite{Choi2019} showed that the perfect or near-perfect fidelity revivals from the $\mathbb{Z}_2$ initial states in the longer ranged deformed models could be explained by an emergent $SU(2)$ algebra in the subspace of the scarred states (whose appropriate raising operators acting on the $\mathbb{Z}_2$ state produces the scarred subspace), with latter work showing multiple such $SU(2)$ algebras corresponding to distinct families of scarred states encompassing the previously known ones \cite{Bull2020} (see also \cite{Omiya2023} for an alternate understanding based on effective spin-1 Hamiltonians). Moreover, the existence of multiple (approximate) $SU(2)$ algebras was argued to govern the short-time and long-time dynamical behaviour of super-diffusive energy transport after appropriately projecting to the constrained Hilbert space in \cite{Ljubotina2023}. Longer ranged generalizations of Eq.(\ref{PXPZ}) exhibit increasingly perfect fidelity revivals and more prominent tower of scar states \cite{Choi2019}, and several other interesting variants and deformations have been considered in the literature, see e.g. \cite{Dooley2020,Ljubotina2023,Chen2023,Shen2024,Kerschbaumer2025}, but we will not consider them in this work and leave them for future studies.

A deformation which has the opposite effect is the following,
\begin{equation}    \label{PXPXP}
    \Delta\mathcal{H}_e = \sum_{i=2}^{N-3} \mathcal{P}_{i-1}\sigma_i^x\mathcal{P}_{i+1}\sigma_{i+2}^x\mathcal{P}_{i+3}  +   \text{   boundary terms              ,}
\end{equation}
which together with the PXP model Eq.(\ref{PXP}) will be referred to as the PXPXP model, $\mathcal{H}_{PXPXP} \!=\! \mathcal{H}_{PXP}+g\Delta\mathcal{H}_e$, where $g$ is the strength of this deformation. It was shown in \cite{Turner2018} that this deformation restores the weakly broken ergodicity of the base PXP model with increasing values of $g$, with $g\!=\!0.25$ identified as the optimal value. Suppression of fidelity revivals and of oscillatory entanglement entropy in favor of a faster (and increasingly linear) growth with increasing values of $g$ is clear in Fig.\ref{fig:fig1} (lower row), signifying washing away of scarring dynamics and gradual restoration of thermalizing dynamics. We will corroborate this also from the point of view of subsystem non-Markovianity, showing that with increasing values of $g$, non-Markovianity of the dynamics of small subsystems diminishes.

We mention at this point that our numerical results have been obtained from second order time-evolving block decimation approach to simulating quantum dynamics \cite{Vidal2004}, using the ITensors.jl library \cite{ITensor}. All results are obtained with open boundary conditions with an optimal timestep size of $\tau\!=\!0.01$, and MPS cutoffs were set at $10^9$, while MPS bond dimensions were not fixed but evolved adaptively with the TEBD flow (however the maximum allowed bond dimension was fixed at 50, which was never attained in our simulations times with the above mentioned MPS cutoff). The subsystems considered are deep in the bulk centered on the middle site, however due to translational invariance their exact location does not matter as long as they are sufficiently away from the boundaries. The total system size considered was a moderate $N\!=\!40$ when initial states are ferromagnetic and Néel states, and $N\!=\!36$ for the 3-CDW initial state (including the results on comparing this initial state with the other two initial states). In Appendix-\ref{appn1}, we present some further numerical details with one example case. More precisely, we show the independence of our results on (i) the measure of distance between quantum states using an alternative measure in Appendix-\ref{appn1a}, (ii) the total system size in Appendix-\ref{appn1b} and (iii) the MPS cutoffs in Appendix-\ref{appn1c}.

\section{Subsystem non-Markovianity}   \label{sec4}

\subsection{Quenching from Néel initial state with PXP and its deformations}

\subsubsection{PXP vs. PXPZ}

This section begins with results on the main theme of this work, that of first demonstrating the non-Markovianity of the dynamics of small subsystems (of a few spins) and then comparing the PXP and PXPZ models with regards to this characteristic. In Fig.\ref{fig:fig2}, we show the non-monotonicities of the trace distances $T_d(\rho_{t+\delta}^{\ell}, \rho_{t}^{\ell})$ (hereafter and in the figures, we shorten our notation and denote this as $T_d(\ell,t,\delta)$) between subsystem states separated in time by $\delta \!=\!1,2,3$ for the PXP (first row) and PXPZ (second row) models. The concerned subsystems comprise of one spin to up to four spins, but the multi-spin subsystems in Fig.\ref{fig:fig2} are not contiguous - any two consecutive spins in the subsystems are separated by one site. That is, for instance, a three-spin subsystem denoted by $\ell\!=\!3$ at the middle of the chain is a subsystem comprising of three spins located at $N/2$, $N/2+2$ and $N/2+4$, and likewise for the two- and four-spin subsystems. In fact, a separation by odd number of sites reproduces the same behaviour (demonstrating this however asks for notably higher numerical costs). When the spins in a subsystem are located at consecutive sites (or they are separated by an even number of sites), remarkably different behaviour (i.e., very diminished) in subsystem non-Markovianity is seen. For two-spin subsystems, this is demonstrated in Appendix-\ref{appn3}. This choice of subsystem configurations is based simply on our numerical observation that such configurations exhibit stronger non-Markovian dynamics, thus we focus on them in the main part of this work. Such subsystem configurations also entail more interfaces between the subsystems in question and their respective environments, which possibly directly are responsible for the enhancement of their non-Markovianity. Additionally, these configurations break the inherent two-site or three-site unit-cell structure respectively of the $\mathbb{Z}_2$ or $\mathbb{Z}_3$ initial states. This facts may also be responsible for the enhanced non-Markovianity of such subsystems, but we leave a detailed investigation of this for future.

Some noteworthy features are revealed immediately from Fig.\ref{fig:fig2}. Firstly, a very regular and orderly behaviour is seen for the case of the PXPZ model compared to that of the PXP model. The oscillatory behaviour is more persistent in the PXPZ model, whereas in the PXP model there is a decaying envelope of the oscillations. Secondly, independent of $\delta$ or the subsystem size, the "deeper" minima of $T_d(\ell,t,\delta)$ are separated by a timescale of $\approx 4.76$ in the PXP model, but in the PXPZ model they are separated by intervals of $\approx 4.52$. At these time intervals, the $\delta$-separated subsystem states come closest to each other, however the extent of their closeness is $\delta$-dependent (in particular, for $\delta\!=\!1$ in the PXPZ model, the trace distance nearly vanishes at these deep minima). This is in line with the periods of the fidelity revivals in Fig.\ref{fig:fig1}, and reveals a rather systematic behaviour of the distances between quantum states (recall that full state fidelity is a unique measure of distances between pure states) across the scales of the spatial extent (full system or small subsystems) of the quantum states. The inverse of the energy gap between the scars in the scarred subspace, which define the time-periods of full-system fidelity revivals, also appear to define the time-periods associated with the temporally-separated subsystem states approaching closest to each other (in terms of the trace distance measure in the Hilbert space). 

It is important to note however that (for a fixed model) only the time periods of the minima at any $\delta$ match with each other and with that of fidelity revivals, but \textit{not} the actual instants of time $t$. Note also that other than the deep minima, milder minima also occur, as seen in Fig.\ref{fig:fig2}. Thirdly, the relative behaviour of individual subsystems compared to the others is dependent on $\delta$ for both PXP and PXPZ cases. For instance, at $\delta\!=\!1$, the larger subsystems exhibit larger levels of non-monotonicities in $T_d(\ell,t,\delta)$, whereas the opposite happens at $\delta\!=\!2$.

\begin{figure*}[tp]
	\centering
        \includegraphics[width=5.6cm,height=5cm]{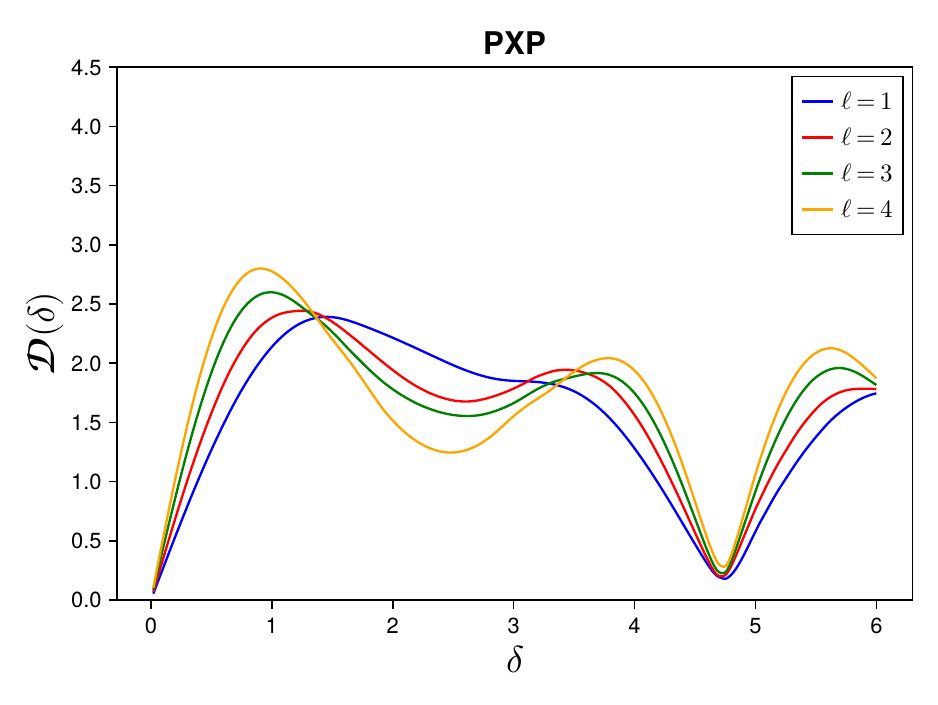}
        \includegraphics[width=5.6cm,height=5cm]{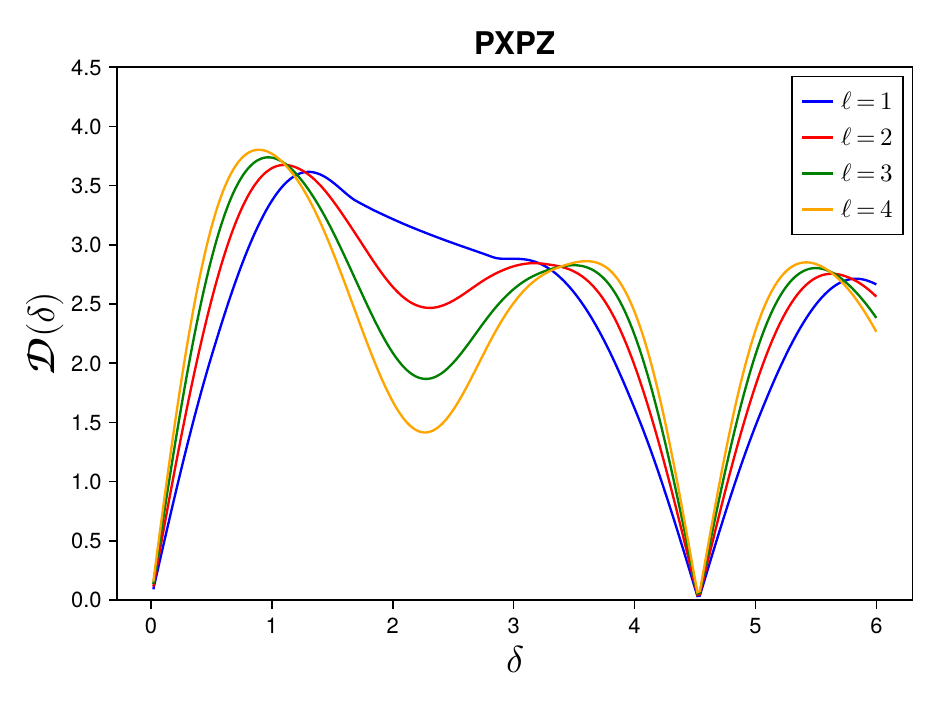}
        \includegraphics[width=5.6cm,height=5cm]{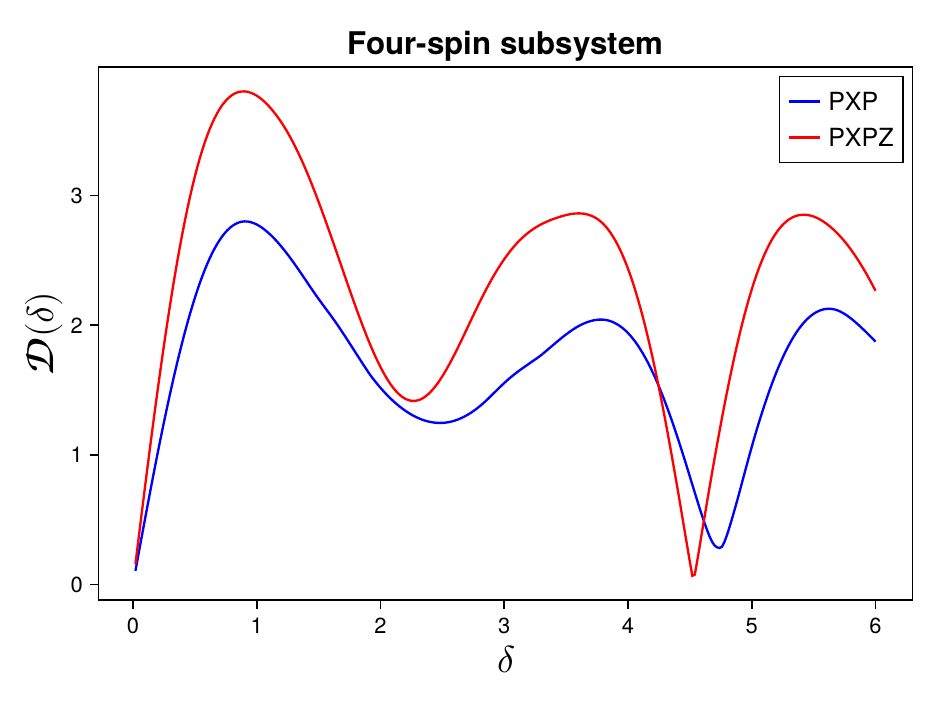}
	\caption{\fontsize{9}{11} \selectfont Subsystem-wise comparisons of the degree of non-Markovianity $\mathcal{D}(\delta)$ (Eq.(\ref{TDdegree})) in the PXP (\textbf{left}) and PXPZ (\textbf{center}) models, and a PXP vs. PXPZ comparison in a four-spin subsystem (\textbf{right}). Global minima of $\mathcal{D}(\delta)$ appears at $\delta\approx 4.76$ in the PXP model and at $\delta\approx 4.52$ in the PXPZ model. Milder minima are also seen for subsystems of size $\ell>1$ at about half these values of $\delta$. No particular pattern is apparent with respect to subsystem sizes, however the PXPZ model exhibits larger $\mathcal{D}(\delta)$ at most values of $\delta$ as apparent in the example of the four-spin subsystem.}
	\label{fig:fig4}
\end{figure*}

\begin{figure*}[!htbp]
	\centering
        \includegraphics[width=5.6cm,height=5cm]{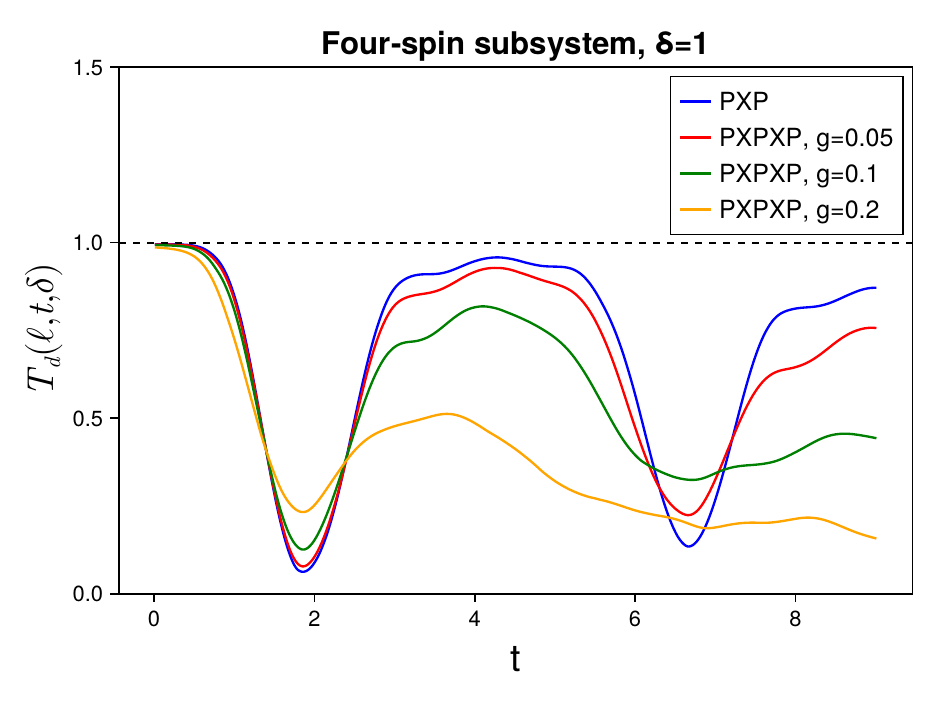}
        \includegraphics[width=5.6cm,height=5cm]{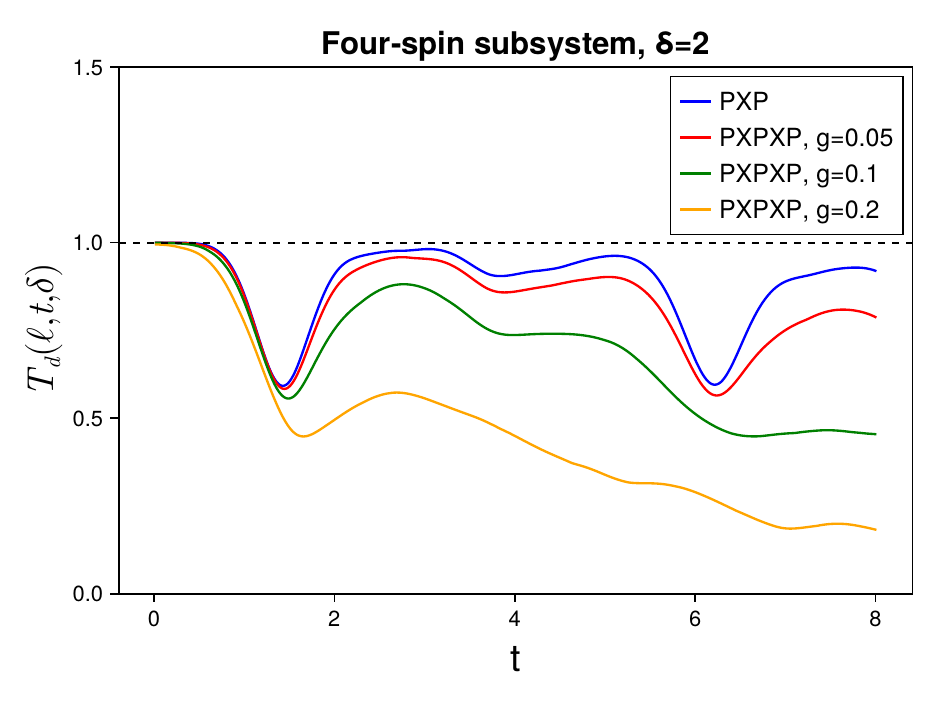}
        \includegraphics[width=5.6cm,height=5cm]{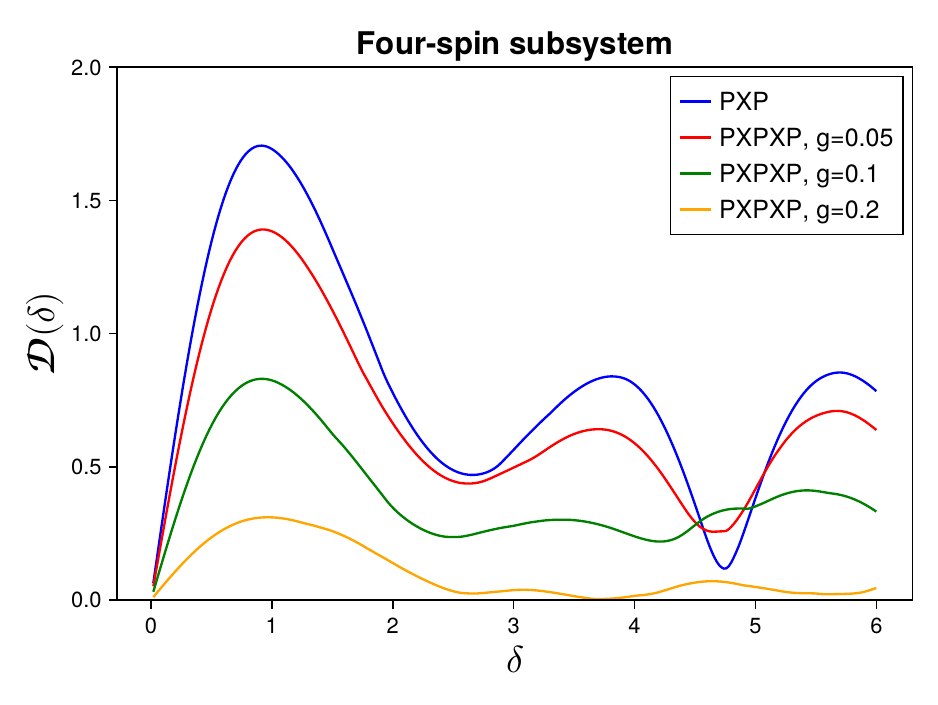}
	\caption{\fontsize{9}{11} \selectfont Ergodicity restoring deformation with increasing strength $g$ in the PXPXP model leads to diminishing non-Markovianity compared to the base PXP model, exemplified with the four-spin subsystem in this figure. Behaviour of the trace distance $T_d(\ell,t,\delta)$ shown for $\delta\!=\!1$ (\textbf{left}) and $\delta\!=\!2$ (\textbf{middle}), and the degree of non-Markovianity $\mathcal{D}(\delta)$ (\textbf{right}).}
	\label{fig:fig5}
\end{figure*}

The degree of non-Markovianity, Eq.(\ref{TDdegree}), displayed in Fig.\ref{fig:fig4} shows no particular pattern with respect to the size $\ell$ of the subsystems (though much more systematic than in the mixed-field Ising spin chain \cite{Banerjee2025a}), except that it attains a global minimum at $\delta \approx 4.76$ for the PXP model and at $\delta \approx 4.52$ for the PXPZ model. Note that these timescales are now with respect to the temporal separation $\delta$. This degree of non-Markovianity is, for a given subsystem, mostly larger in the case of PXPZ model compared to that in the PXP model for most considered values of the temporal separation $\delta$, as was also evident for the three specific values of $\delta$ in Fig.\ref{fig:fig2}. Moreover, in the case of the PXPZ model, the global minima in $\mathcal{D}(\delta)$ is attained more sharply and at a value $\approx 0$. In Appendix-\ref{appn4}, we show that a similar degree enumerating the increasing-ness of the "classical" counterpart of trace distance between the eigenvalues of the reduced density matrices corresponding to $\delta$-separated subsystems shows a more systematic behaviour with regards to the size of the subsystems, and the timescales in $\delta$ above appear there as well.

\subsubsection{PXP vs. PXPXP}

We now compare subsystem non-Markovianity between the PXP and PXPXP models. Since the PXPXP deformation tries to reinstate ergodicity by having a destructive effect on the scarred subspace \cite{Turner2018}, thereby enhancing the approach to thermalization, and since non-Markovianity in general prohibits an efficient relaxation to the equilibrated state (in this case, thermalized state), it is to be expected that this deformation gradually weakens any signatures of non-Markovianity of subsystem dynamics. This is corroborated in Fig.\ref{fig:fig5} for increasing values of the parameter $g$, where we show the results only for the four-spin subsystem for clarity. For higher values of $g$, the non-monotonicities of the trace distances $T_d(\ell,t,\delta)$ gradually diminish for a given $\delta$ as seen in the left and middle figures in Fig.\ref{fig:fig5}. Consequently, the degree of non-Markovianity $\mathcal{D}(\delta)$ decreases and in fact for $g\!=\!0.2$ is very low and hovers near zero for $ \delta \gtrsim 2.5$ (that is, dynamics gradually becomes effectively Markovian in the sense of weakening or no information backflows).

A remark is due at this point. While the relationship between ergodicity and Markovianity is more subtle and not fully understood in general, it is expected that a dynamical process in which the exploration of the full available state space happens "efficiently" and without prohibition should entail Markovianity (memory-less-ness) of the underlying dynamics, although the converse need not necessarily be true (for instance, a dynamics restricted within a subspace may also be Markovian). It is plausible that Markovian dynamics implies ergodicity when the former is established across a broad range of initial states, in other words when Markovianity ensues from \textit{typical} initial states. Recall that even scarred quantum systems are expected to exhibit ergodic dynamics when initiated from generic, typical initial states that have little support on the scarred subspace but are instead well-spread over many eigenstates of the underlying Hamiltonian. It is only for atypical initial states that are mostly or entirely supported on the scarred subspace that the ensuing dynamics breaks ergodicity (at least in a weak sense). Be that as it may, in the particular case of the PXPXP model, as the strength of the deformation increases and model grows closer to restoring ergodicity, dynamics of small subsystems increasingly approaches Markovianity (as defined by the lack of information backflows).

\subsection{PXP quench from $\mathbb{Z}_3$ initial state}\label{z3}

\begin{figure}[!htbp]
	\centering
        \includegraphics[width=6.4cm,height=6cm]{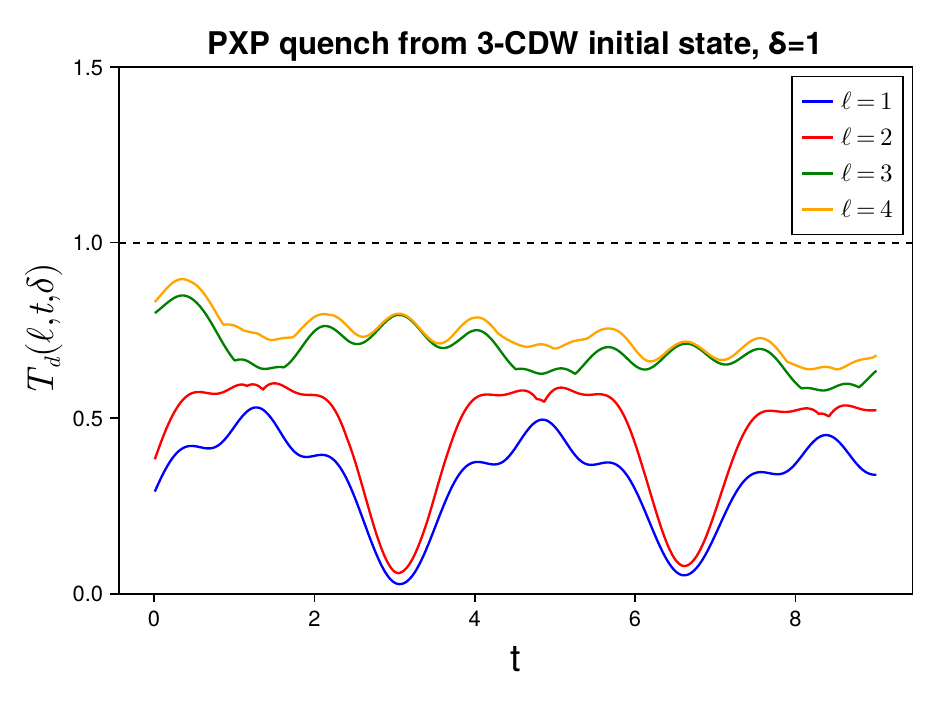}
        \includegraphics[width=6.4cm,height=6cm]{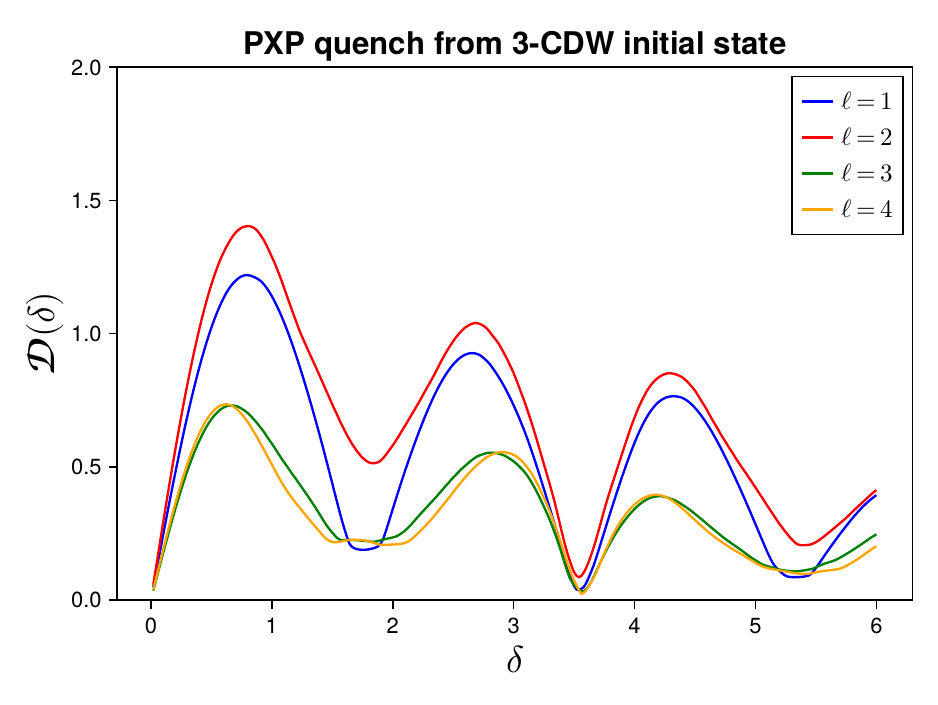}
	\caption{\fontsize{9}{11} \selectfont For the PXP quench from the 3-CDW ($\mathbb{Z}_3$) initial state, shown here are subsystem-wise comparisons of the trace distance $T_d(\ell,t,\delta)$ for the example of $\delta\!=\!1$ (\textbf{upper}) and degree of non-Markovianity $\mathcal{D}(\delta)$ \textbf{lower}. In contrast to the case of PXP quench from the Néel ($\mathbb{Z}_2$) initial state displayed in Fig.\ref{fig:fig4}, the dynamics of the smaller subsystems made of one- and two-spins is more non-Markovian than that of three- and four-spin subsystems.}
	\label{fig:figZ3}
\end{figure}

 For the case of PXP quench initiated from the $\mathbb{Z}_3$ initial state, in Fig.\ref{fig:figZ3} we show a subsystem-wise comparisons of the behaviour of the trace distance $T_d(\ell,t,\delta)$ for the case of $\delta\!=\!1$, as well as the behaviour of the non-Markovianity degree $\mathcal{D}(\delta)$. It is interesting to note here that in this case, the dynamics of one- and two-spin subsystems is notably more non-Markovian than that of three- and four-spin subsystems. This is in contrast to the case of the PXP quench from the Néel ($\mathbb{Z}_2$) initial state displayed in Fig.\ref{fig:fig4} (left), where all the considered subsystems exhibit fairly comparable levels of non-Markovianity.

\subsection{Average non-Markovianity degree} \label{avgdegree}

\begin{figure}[!htbp]
	\centering
        \includegraphics[width=6.4cm,height=6cm]{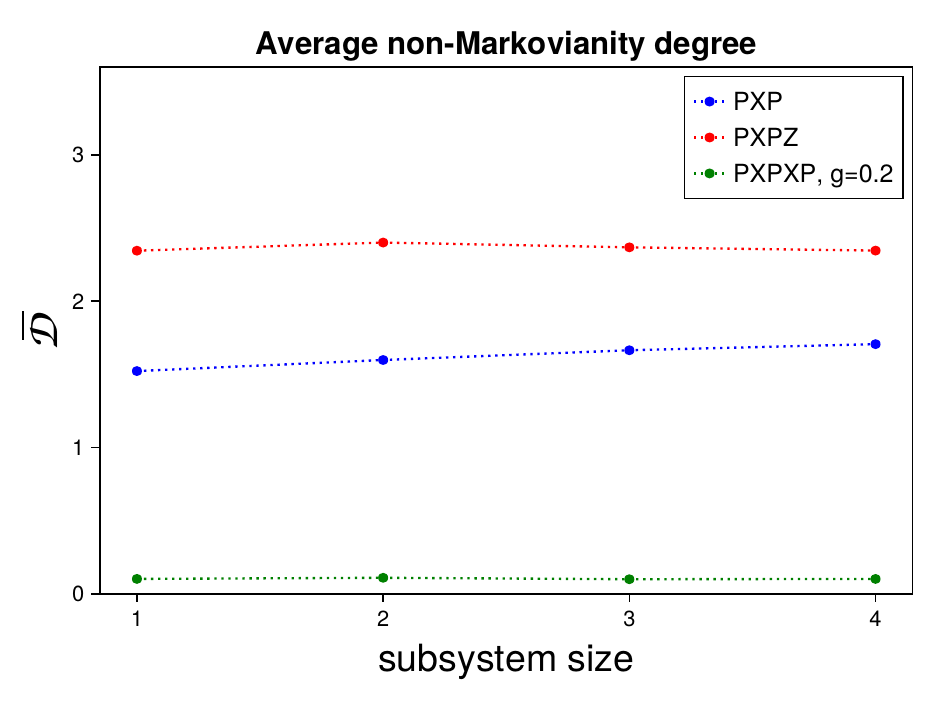}
        \includegraphics[width=6.4cm,height=6cm]{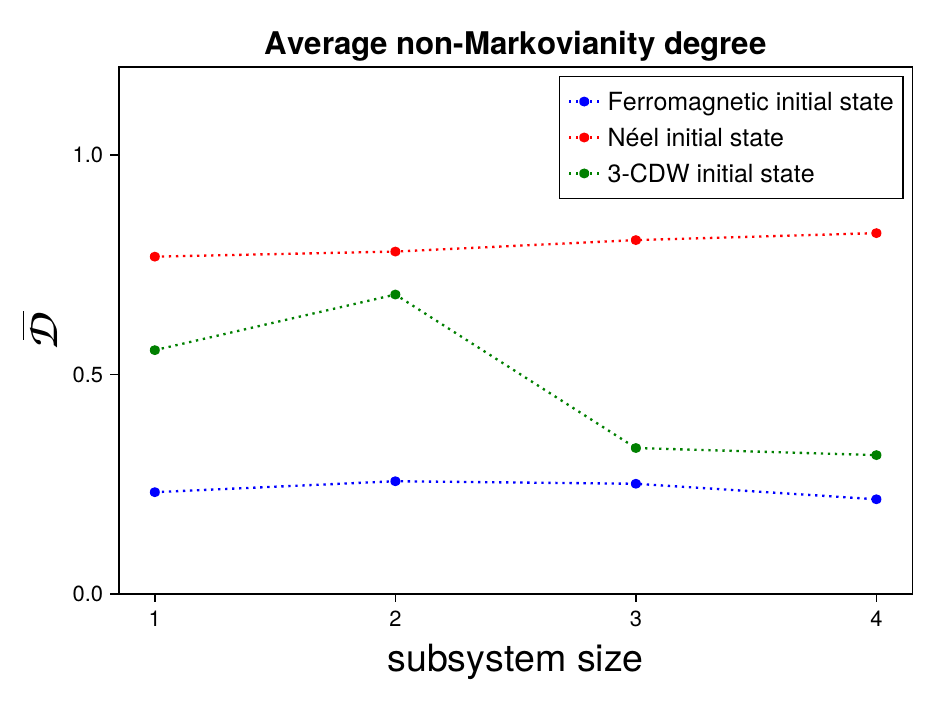}
        
	\caption{\fontsize{9}{11} \selectfont Average non-Markovianity degree defined in Eq.(\ref{avgdegNM} versus subsystem size. In the (\textbf{upper}) figure is shown model-wise comparison with the Néel initial state, and in the (\textbf{lower}) figure is shown initial state-wise comparison for the PXP model. Average degree of subsystem non-Markovianity unambiguously shows enhancement (or diminishment) of subsystem non-Markovianity is concurrent with enhancement (or diminishment) of quantum many-body scarring. Note that in the upper figure of model-wise comparisons, the non-Markovianity degrees $\mathcal{D}(\delta)$ were collected from trace distance revivals over simulations times up to $t\!=\!20$, whereas in the lower figure of initial state-wise comparisons, the non-Markovianity degrees were collected from trace distance revivals over simulations times up to $t\!=\!10$. This leads to larger numerical values of the the non-Markovianity degree and its average in the former case.}
	\label{fig:avgNMdeg}
\end{figure}

The above results on the degree of non-Markovianity $\mathcal{D}(\delta)$ in Figs.(\ref{fig:fig4}, \ref{fig:fig5}, \ref{fig:figZ3}) reveal a deficiency in its definition Eq.(\ref{TDdegree}). As seen in these figures, for certain values of the temporal separation $\delta$, the curves of the non-Markovianity degree cross each other in the case of both subsystem-wise comparisons (revealing that subsystem sizes play lesser role, as remarked earlier) and more importantly in the case of deformation-wise comparisons (right-most figures in Figs.(\ref{fig:fig4}, \ref{fig:fig5}). In the latter case, it is seen that for certain small windows of $\delta$, the scars enhancing (diminishing) deformations may show reduced (increased) subsystem non-Markovianity compared to the PXP model. Thus, while the definition in Eq.(\ref{TDdegree}) correctly captures the overall tendencies of scars enhancing or diminishing deformations with regards to subsystem non-Markovianity, it nonetheless may lead to opposite conclusions if one were to observe only the aforementioned small windows of $\delta$.

To rectify this deficiency and make our conclusions more precise, one may consider a maximization over $\delta$ in Eq.(\ref{TDdegree}) following the approach of Ref.(\cite{Breuer2009}). However, this will display the degrees of non-Markovianity only at the specific value of $\delta$ that maximizes this degree for a given case, without necessarily addressing the aforementioned deficiency. Moreover, the maxima of $\mathcal{D}(\delta)$ occur around $\delta_{max}\!\sim\!1$ in all cases considered, and the resultant trace distance dynamics are similar to the previously shown instances at $\delta\!=\!1$.

We therefore address this issue by considering the average degree of non-Markovianity, 
\begin{equation}   \label{avgdegNM}
    \overline{\mathcal{D}} \!=\! \sum_{\delta_i} \mathcal{D}(\delta_i) / S , 
\end{equation}
where the average is taken over the various values of temporal separations $\delta_i$ considered in our simulations and $S$ denotes the number of elements in the set $\{\delta_i\}$. This is shown in Fig.\ref{fig:avgNMdeg}, for both the cases of PXP and its deformations with the Néel initial state as well as PXP quench initiated from the three different initial states. We see that the average non-Markovianity degree unambiguously shows enhancement (or diminishment) of subsystem non-Markovianity with enhancement (or diminishment) of quantum many-body scarring. We also note that $\overline{\mathcal{D}}$ is only mildly dependent on subsystem size for the range of subsystems considered in this work, except for the case of the initial state being the 3-CDW state where the smaller subsystems of one and two spins shown noticeably larger non-Markovianity that the larger subsystems.

\subsection{Comparing PXP quench from different initial states}\label{z1vsz2vsz3}

In this subsection, we compare subsystem non-Markovianity associated to the dynamics induced by quenching with the PXP Hamiltonian with three different initial states mentioned previously - the thermalizing fully polarized ferromagnetic state, the Néel ($\mathbb{Z}_2$) state exhibiting scarring dynamics, and the 3-CDW ($\mathbb{Z}_3$) state which exhibits a much weakened scarring dynamics. This is shown in Fig.\ref{fig:avgNMdeg} (lower), where we have compared the average degree of non-Markovianity $\mathcal{D}(\delta)$ (collected over dynamics up to time $t\!=\!10$) resulting from the three initial states. Expectedly, the subsystem non-Markovianity is much lower (though not zero within our simulation times) when the initial state is the ferromagnetic one.

\section{Conclusion}     \label{sec5}

In this work, by invoking notions from the domain of open quantum systems and exemplified with the PXP model and its relevant deformations, we have numerically uncovered a close relationship between quantum many-body scarring in closed quantum many-body systems (which show unfrozen entanglement dynamics) and non-Markovian dynamics of subsystems of such systems. While the retention (or a very slow loss) of the memory of the initial state is generic (and obvious) in scarred quantum dynamics, here we have studied non-Markovianity of subsystem dynamics as quantified by the violations of contractivity of the (trace) distances between temporally-separated transient states of a given subsystem \cite{Banerjee2025a} to reveal a finer form of memory retention associated with scarred quantum dynamics.

We have found that the PXPZ-type deformations of the PXP model (with Néel initial state) which stabilize or enhance quantum scarring also stabilize or enhance subsystem non-Markovianity, while the PXPXP-type deformations which wash away quantum scarring also have the same effect on subsystem non-Markovianity. Scars-enhancing deformations lead to stronger enforcement of subspace-restricted dynamics (significantly disconnected from the rest of the spectrum), and this leads to retention of memory between transient subsystem states, which is a finer form of memory effects than captured by full system fidelity revivals. Moreover, the time-periods of revivals of fidelity with the initial state also sets other time-periods associated with subsystem non-Markovianity discussed in the previous sections. Likewise, we have also seen that on varying the initial state on which to enforce a PXP quench from thermalizing (ferromagnetic) initial states to most non-thermalizing (Néel) initial states, subsystem non-Markovianity becomes stronger. These results are essentially consequences of global state revivals and long-lived oscillations exhibited by such non-thermalizing systems, and the perspective of subsystem non-Markovianity is a previously unexplored attribute that provides an additional, if not an alternative, point of view towards such systems and their dynamics.

We expect these features to hold more broadly for other scars-enhancing (or -diminishing) deformations of the PXP model, and more generally in other scarred quantum systems which show unfrozen/oscillatory entanglement dynamics when initiated from initial states with large support on their respective scarred subspaces. Indeed, even in the case of the mixed-field Ising spin chain studied in Ref.(\cite{Banerjee2025a}), a less systematic connection between scars and subsystem non-Markovianity was somewhat apparent. This is because in that work, strong subsystem non-Markovianity was seen only in the quenching protocol (paramagnetic-to-magnetic) with strong confinement amongst quasiparticles, and confinement has been known to generically lead to scars-type non-thermal eigenstates \cite{James2019} (and confinement-induced constraints on the propagation of excitations generically leading to slow thermalization and a prolonged pre-thermalization regime, see e.g. \cite{Kormos2016,Birnkammer2022}).

Note that we have not yet declared that the memory effects underlying subsystem non-Markovianity are "purely" quantum in nature, as this is a very subtle matter in general and requires separate care, see e.g. \cite{Giarmatzi2021,Romero2019,Milz2020,Banacki2023,Backer2024,Budini2025,Backer2025,Bravo2025}. Indeed, in Appendix-\ref{appn4}, we see systematic non-monotonicities in a "classical" counterpart of the trace distance measure. It would be very interesting in future to clarify this and the more general issue of classifying and quantifying purely-quantum memories in the subsystems' dynamics.

This work has provided numerical evidence that enhanced scarring leads to enhanced subsystem non-Markovianity, but does the converse hold true at all ? It is well appreciated by now that the presence of quantum scars (and initial states with support on them) leads to many-body revivals, but a stronger (and the converse) case has been argued that revivals imply the existence of quantum scars \cite{Alhambra2020}. It is intriguing to wonder if arguments in the spirit of \cite{Alhambra2020} can be used to shed light on the question of whether persistent non-Markovianity of subsystem dynamics also implies the existence of quantum many-body scars.

Aside from the non-thermal scar states arising out of kinetic constraints in quantum many-body systems such as the PXP family of models considered in this work, a more traditional notion of scarring in quantum many-body systems, which is not dependent on the existence of any constraints in the Hilbert space, has recently been proposed based on generalizing the scarring (in classical phase space) of individual eigenstates in single-particle quantum chaotic systems and therefore can be considered as "genuine" scarring \cite{Evrard2024,Pizzi2024,Ermakov2024}. These can be both thermal (in the sense of satisfying ETH) or non-thermal depending on whether they are associated with unstable or stable periodic orbits in the corresponding classical phase spaces \cite{Evrard2024}. These types of scars also give rise to memory effects \cite{Pizzi2024}, but the latter can exist in weaker form even without such scars \cite{Pizzi2025,Graf2024}. It would be very interesting to extend and apply our notions of non-Markovianity in this setting of genuine scarring in classical phase spaces associated with quantum many-body systems.

Many other measures and notions of non-Markovianity exist in the open quantum systems literature. While most information backflows-based measures are expected to yield qualitatively similar results, it would be very interesting to investigate subsystem non-Markovianity in scarred and other closed quantum many-body systems  using approaches based on process tensors or properties of the extracted dynamical maps underlying subsystem dynamics, see e.g. \cite{Pollock2018,Pollock2018a,Guo2020,Guo2022,Cerrillo2014,Rosenbach2016,Strachan2024,Coppola2025}). Separately, it would also be very interesting to see if the strongly non-Markovian dynamics of subsystems seen in this work (together with the differences between PXP and PXPZ in this context) can be emulated and further elucidated with the collision models framework \cite{Ciccarello2022}.

\section*{Acknowledgments}

Financial support in the form of a postdoctoral fellowship from the Department of Atomic Energy, India is gratefully acknowledged. We also acknowledge the support from CQuERE, TCG CREST, India during the finishing stages of this work. Additionally, we thank an anonymous referee for useful comments which helped solidify the main message of this work.

\section*{Data availability}

All numerical data on which this work is based can be obtained from the author on reasonable request.

\appendix

\section{Further numerical details}
\label{appn1}

\subsection{Subsystem non-Markovianity via fidelity measure}
\label{appn1a}

\begin{figure}[!htbp]
	\centering
	\includegraphics[width=4.25cm,height=4.2cm]{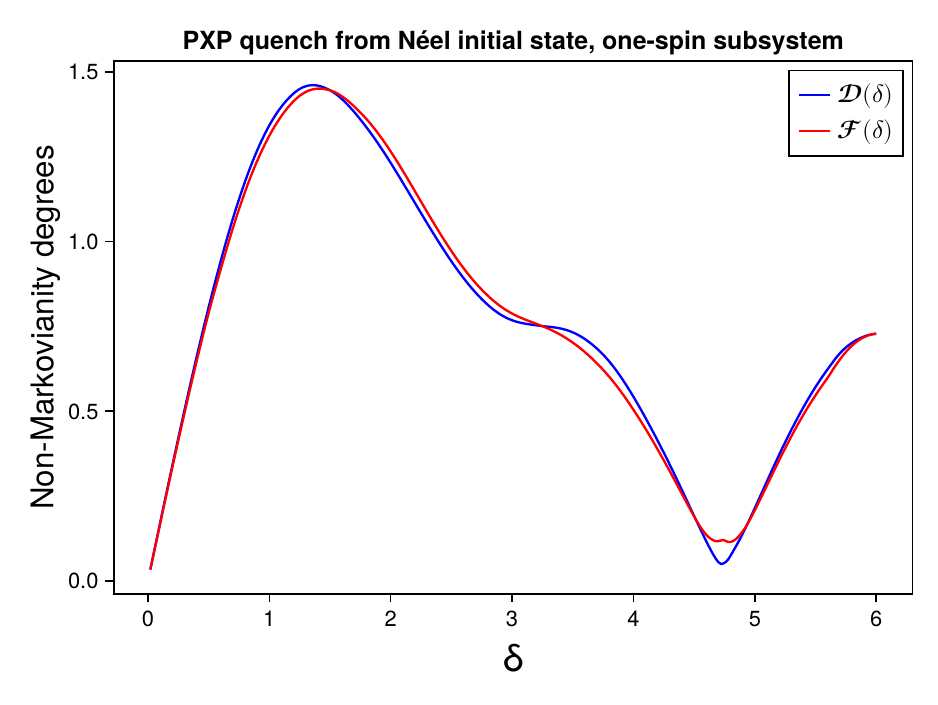}
    \includegraphics[width=4.25cm,height=4.2cm]{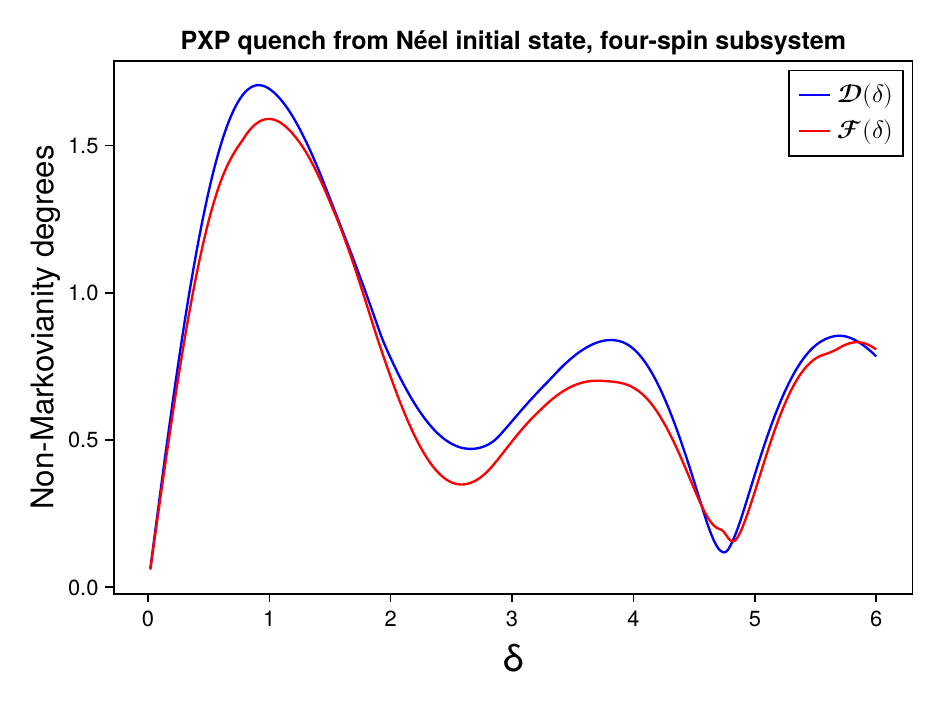}
	\caption{\fontsize{9}{11} \selectfont Qualitative agreement for subsystem non-Markovianity as measured by the trace distance and fidelity based measures $\mathcal{D}(\delta)$ and $\mathcal{F}(\delta)$ respectively, shown here for the examples of the non-Markovianity degree $\mathcal{D}(\delta)$ of one- and four-spin subsystems in the case of PXP quench from Néel initial state.}
	\label{fig:TDvsFID}
\end{figure}

In the main text, we mentioned that our results are qualitatively unchanged if measures other than the trace distance (between temporally-separated subsystem states) are chosen. Here we provide two samples of the evidence towards that claim, by quantifying subsystem non-Markovianity using fidelity between temporally-separated subsystem states. 

The fidelity between temporally-separated subsystem states $\rho^{\ell}_{t} $ and $\rho^{\ell}_{t+\delta} $ is given by \cite{Jozsa1994},
\begin{equation}   \label{fidelity}
    F(t,\delta) = \operatorname{tr} \sqrt{\sqrt{\rho^{\ell}_{t+\delta}}\text{   }\rho^{\ell}_{t} \text{   }\sqrt{\rho^{\ell}_{t+\delta}}} .
\end{equation}

While the fidelity is not a proper distance measure by itself, a useful distance measure built of it is the Bures distance, $B(t,\delta) \!=\! \sqrt{1 - F(t,\delta)^2}$ \cite{Hayashi2017}. Advocating to the contractivity of the Bures distance under CPTP maps, a useful measure of non-Markovianity may be defined by enumerating its violation of this contractivity \cite{Liu2013}, which for our case reads, 

\begin{equation}     \label{Buresdegree}
    \mathcal{F}(\delta) = \sum_t \alpha_2(t,\delta)  \hspace{0.3cm} \forall t \hspace{0.25cm} \text{s.t.} \hspace{0.25cm} \alpha_2(t,\delta)>0    \text{ ,}
\end{equation}
where,
\begin{equation}    \label{slope2}
    \alpha_2(t,\delta) = \frac{1}{\tau}\bigg(B(t+\tau,\delta) - B(t,\delta) \bigg)   \text{    .}
\end{equation}

In Fig.\ref{fig:TDvsFID}, we show a comparison between the non-Markovianity degrees $\mathcal{D}(\delta)$ and $\mathcal{F}(\delta)$ for one- and four-spin subsystems in the case of the PXP quench from the Néel initial state. Qualitative agreement (and in fact, close quantitative agreement as well) between the two measures is clearly evident. This holds also for all other cases considered in this work.

\subsection{System size independence}
\label{appn1b}

\begin{figure}[!htbp]
	\centering
	\includegraphics[width=4.25cm,height=4.2cm]{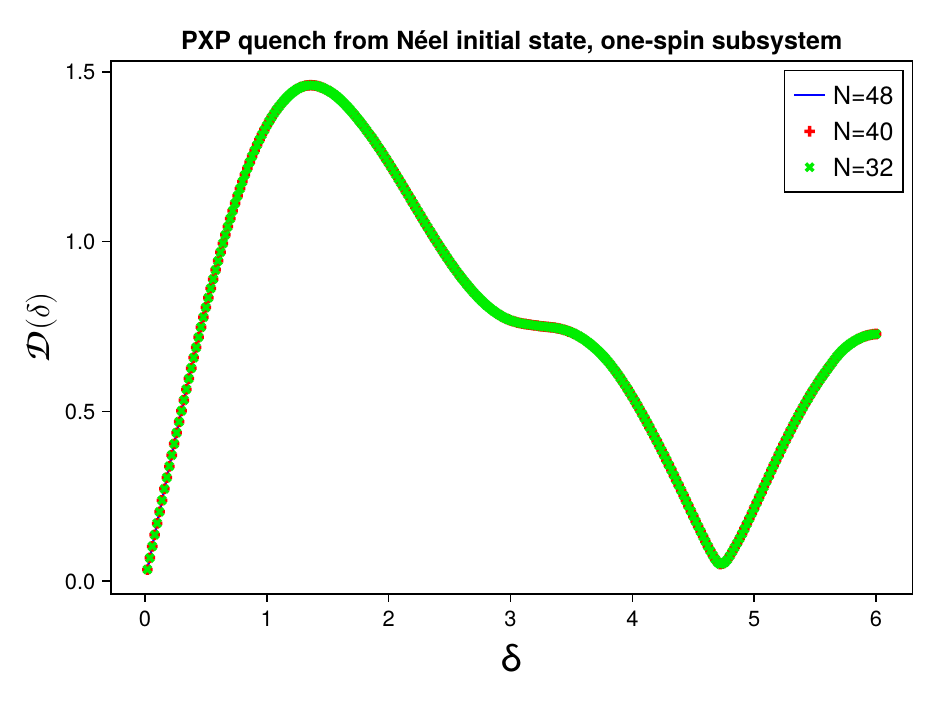}
    \includegraphics[width=4.25cm,height=4.2cm]{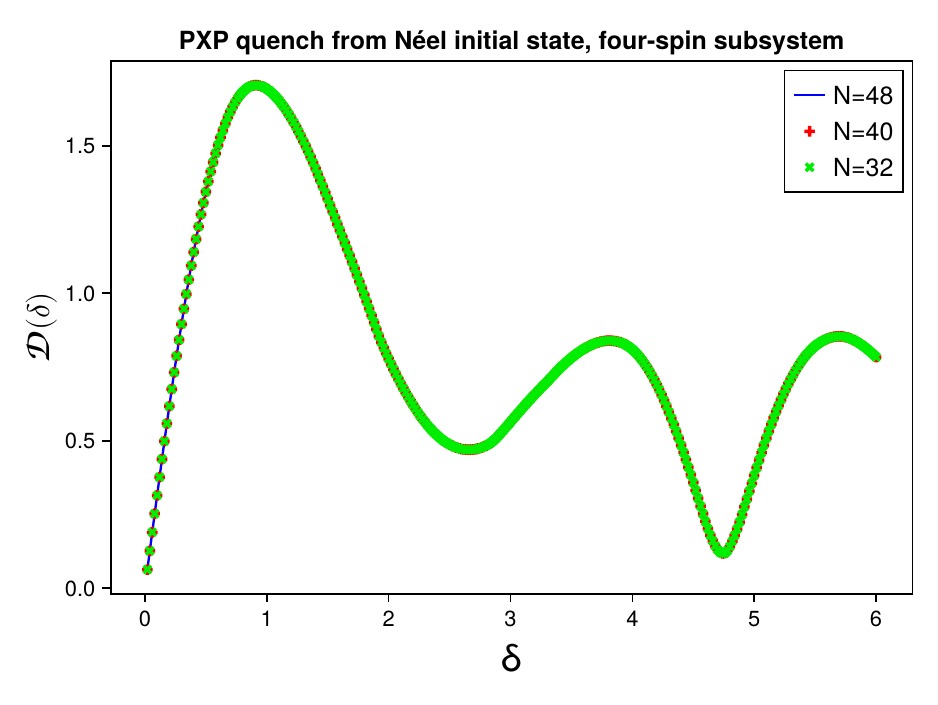}
	\caption{\fontsize{9}{11} \selectfont Full independence on total system sizes of our results on subsystem non-Markovianity, shown here for the examples of the non-Markovianity degree $\mathcal{D}(\delta)$ of one- and four-spin subsystems in the case of PXP quench from Néel initial state.}
	\label{fig:systemsize}
\end{figure}

As a sample of system size independence of our results on subsystem non-Markovianity, in Fig.\ref{fig:systemsize} we show a system size comparison (for three system sizes $N\!=\!\{32,40,48\}$ of the degree of non-Markovianity $\mathcal{D}(\delta)$ for one- and four-spin subsystems in the case of quenching the Néel initial state by the PXP Hamiltonian. Such system size independence holds for all other cases as well. This is expected due to the subsystems being much smaller compared to the full system sizes and situated far away from the boundaries, as well as due to the non-thermalizing nature of the system's dynamics at large. 

\subsection{MPS cutoff dependence}
\label{appn1c}

\begin{figure}[!htbp]
	\centering
	\includegraphics[width=4.25cm,height=4.2cm]{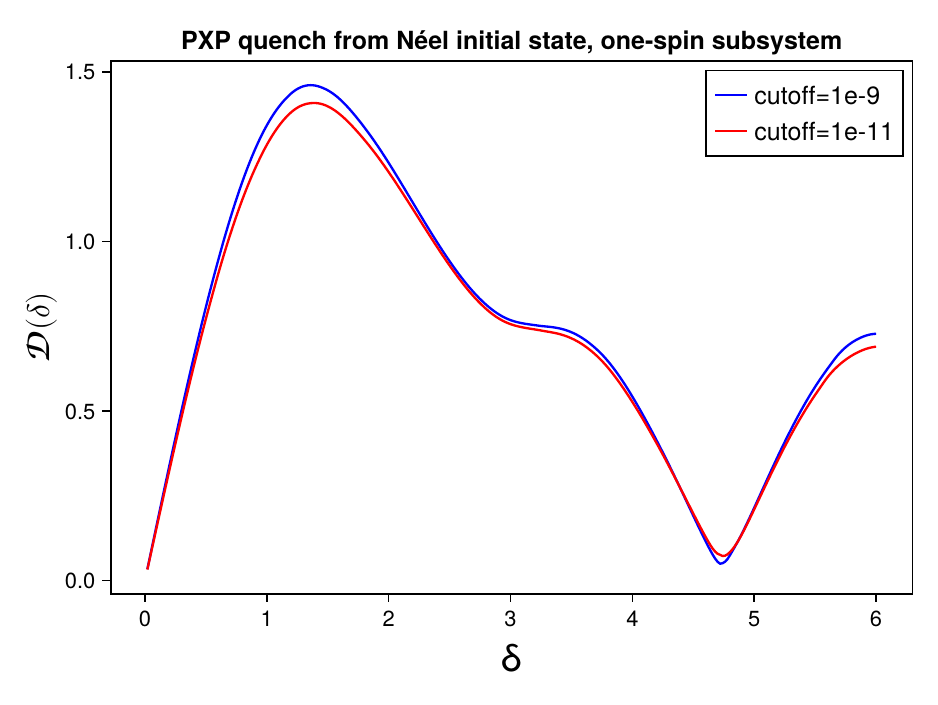}
    \includegraphics[width=4.25cm,height=4.2cm]{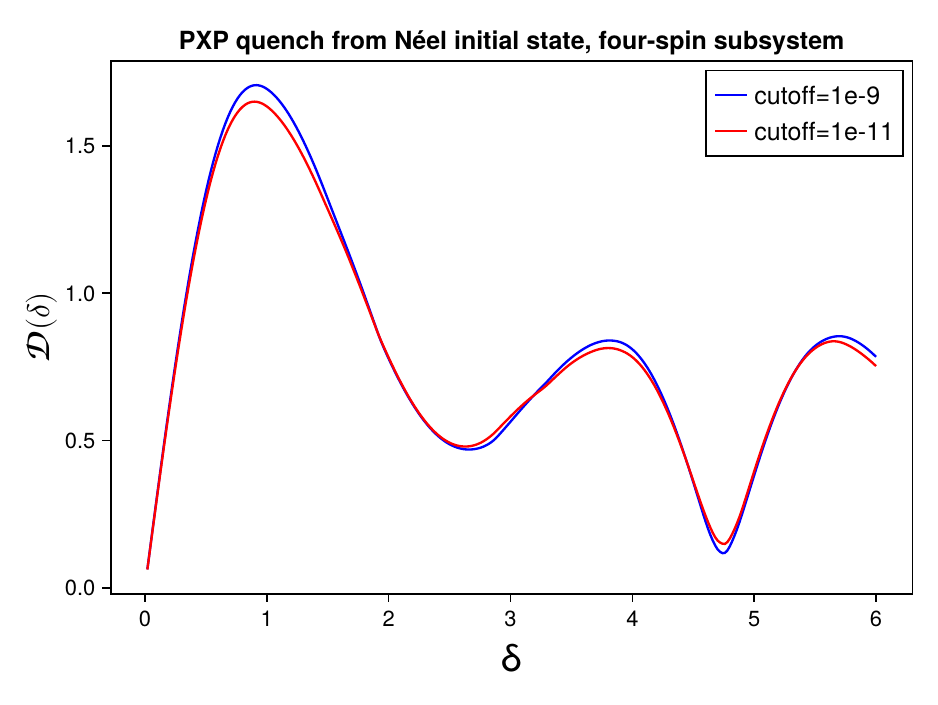}
	\caption{\fontsize{9}{11} \selectfont Qualitative and almost-quantitative independence of our results with respect to the numerical MPS cutoffs, shown here for the examples of the non-Markovianity degree $\mathcal{D}(\delta)$ of one- and four-spin subsystems in the case of PXP quench from Néel initial state. }
	\label{fig:cutoff}
\end{figure}

In Fig.\ref{fig:cutoff}, for the same case as in the previous subsection, we show the qualitative and almost-quantitative indifference of the non-Markovianity degree on the MPS cutoffs enforced in our simulation. This holds also for all other cases considered in this work.

\begin{figure*}[!htbp]
	\centering
        \includegraphics[width=4.4cm,height=4.2cm]{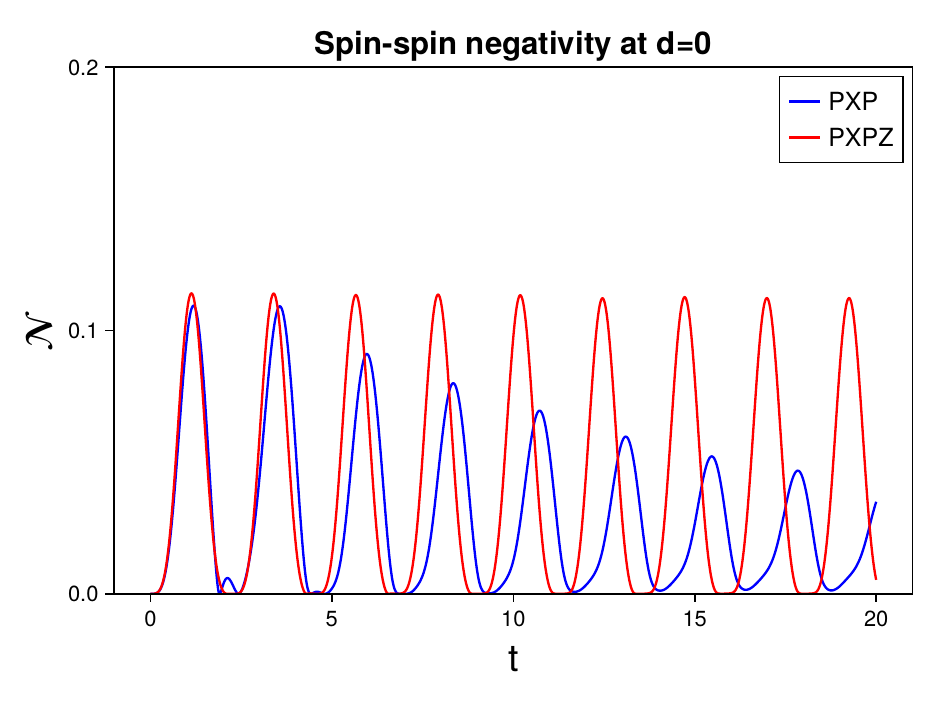}
        \includegraphics[width=4.4cm,height=4.2cm]{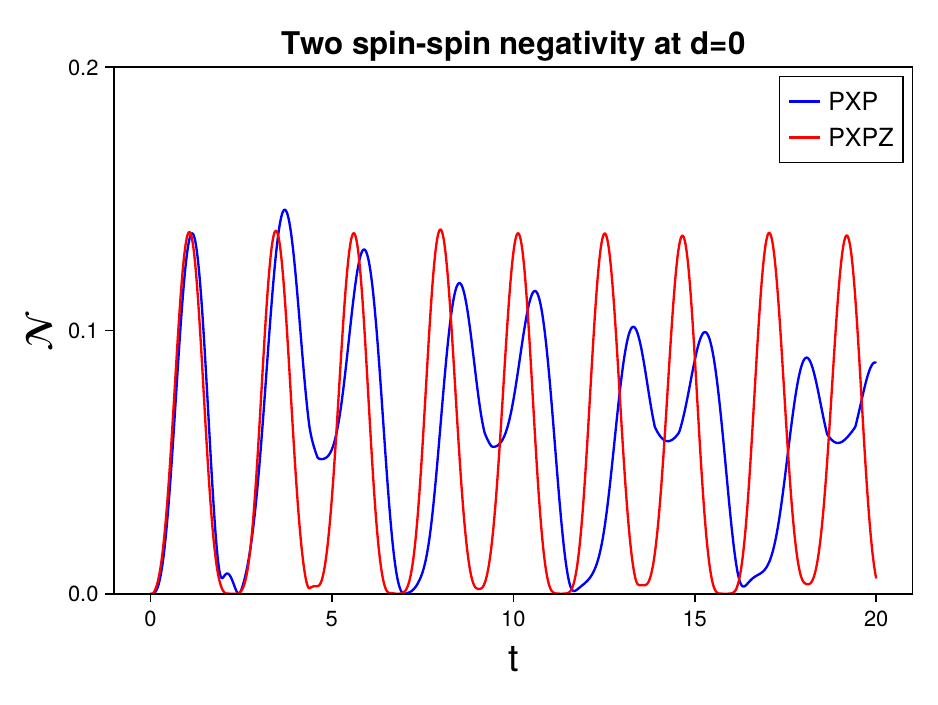}
        \includegraphics[width=4.4cm,height=4.2cm]{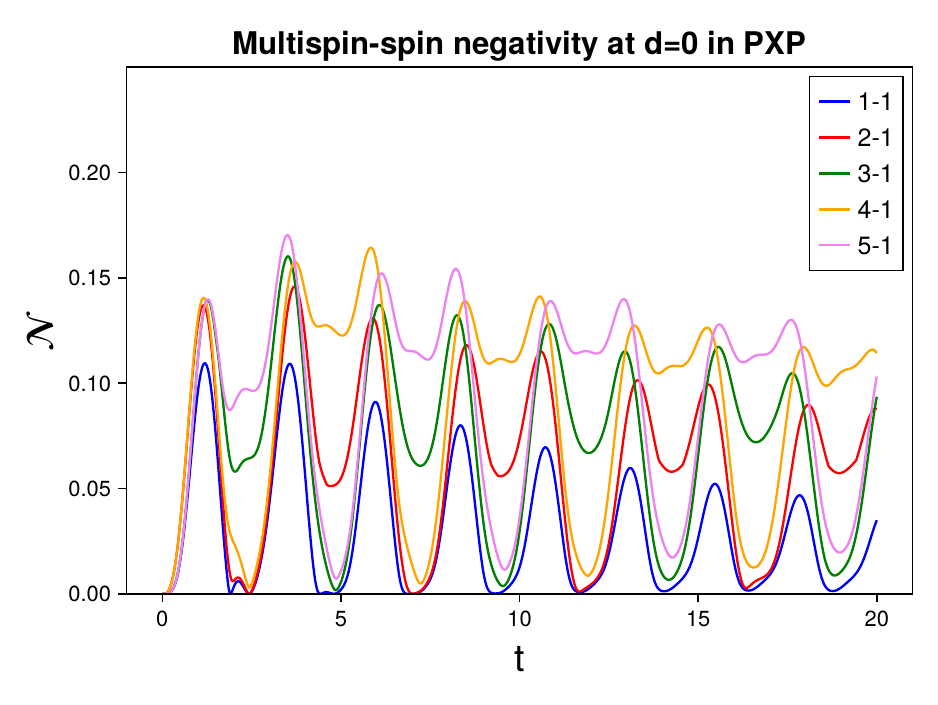}
        \includegraphics[width=4.4cm,height=4.2cm]{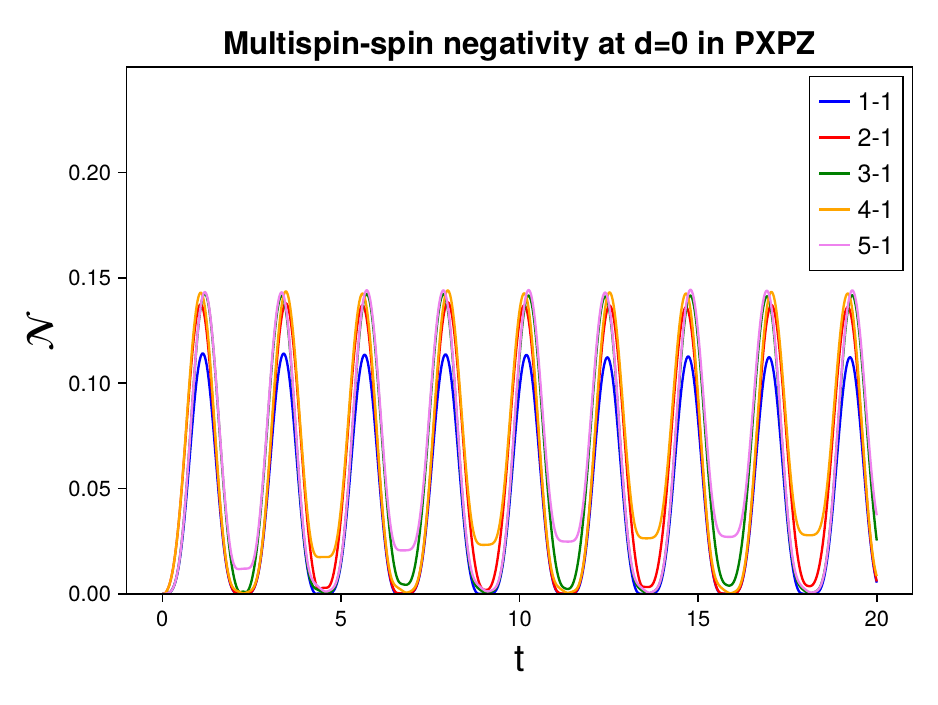}
	\caption{\fontsize{9}{11} \selectfont Dynamics of negativity, Eq.(\ref{neg}), between adjacent (separation $d\!=\!0)$ subsystems, where the transposed subsystem is always a single spin for simplicity. In the labels of the third and fourth figures, $k-1$ denotes the negativity between a block of $k \!=\!\{1,2,3,4,5\}$ contiguous spins and an adjacent spin. In all cases, the negativity oscillates persistently in PXPZ model, while a decaying envelope is seen for the PXP case.}
	\label{fig:fig6}
\end{figure*}

\section{Negativity dynamics}              \label{appn2}

Here we present some auxiliary results on the dynamics of mixed-state bipartite entanglement between spins in small subsystems embedded in the bulk of the system. Since such spins are in mixed states (being reduced from the full system's pure state), we use the most popular and an easily computable measure of mixed state bipartite entanglement called the negativity of entanglement. Given a quantum state $\rho_{AB}$ which exists in a joint Hilbert space $\mathcal{H}_A \otimes \mathcal{H}_B$ of  subsystems $A$ and $B$, one considers its partial transpose with respect to one of the subsystems, say B, denoted by $\rho_{AB}^{T_B}$. The positive partial transpose (PPT) criterion \cite{Peres1996,Horodeckis1996} declares that if $\rho_{AB}$ is separable, then $\rho_{AB}^{T_B}$ is also a density matrix, i.e., $\rho_{AB}^{T_B} \in \mathcal{D}(\mathcal{H}$) . Consequently, if $\rho_{AB}^{T_B}$ has any negative eigenvalues, then it is not a physical state, i.e., $\rho_{AB}^{T_B} \notin \mathcal{D}(\mathcal{H}$). To quantify this entanglement, one defines entanglement negativity of $\rho_{AB}$ in terms of the eigenvalues $\{p_j\} $ of $\rho_{AB}^{T_B}$ \cite{VidalWerner2002}, 
\begin{equation}   \label{neg}
    \mathcal{N}(\rho_{AB}) = \frac{1}{2} \underset{j}{\sum} \big( |p_j| - p_j \big)  \text{      ,}
\end{equation}
which clearly just counts the total magnitude of the negative eigenvalues.

Let us mention that the PPT criterion is only a necessary condition for separability, but for qubit-qubit (Hilbert space dimension $2 \otimes 2$) or qubit-qutrit (Hilbert space dimension $2 \otimes 3$) systems, it is also known to be \textit{sufficient} \cite{Horodeckis1996}. Nonetheless, existence of any negative eigenvalues of $\rho_{AB}^{T_B}$ guarantees that the constituents $A$ and $B$ of the system $AB$ are entangled with each other (at least as detected by the partial transposition operation).

The first and second figures in Fig.\ref{fig:fig6} respectively show the comparative (PXP vs. PXPZ) dynamics of negativity between the two nearest neighbor spins, and between a two-adjacent-spins subsystem with a third spin adjacent to it (separation $d\!=\!0$). In both cases, the PXPZ model shows persistent oscillations of the negativity, while it decays for the PXP model.

The third and fourth figures in Fig.\ref{fig:fig6} show a subsystem size comparisons (for each of PXP and PXPZ models separately) of the negativity between a contiguous block of $k$ spins (where $k\!=\!\{1,2,3,4,5\}$ with a spin adjacent to it (separation $d\!=\!0$). The PXPZ model exhibits cleaner and persistent oscillations (which also mostly overlap with each other) of the negativity dynamics of the considered subsystems, whereas in the PXP model, the negativity oscillations are less systematic and exhibit a slowly decaying profile of the maxima of the oscillations.

\begin{figure*}[!htbp]
	\centering
        \includegraphics[width=5.6cm,height=5cm]{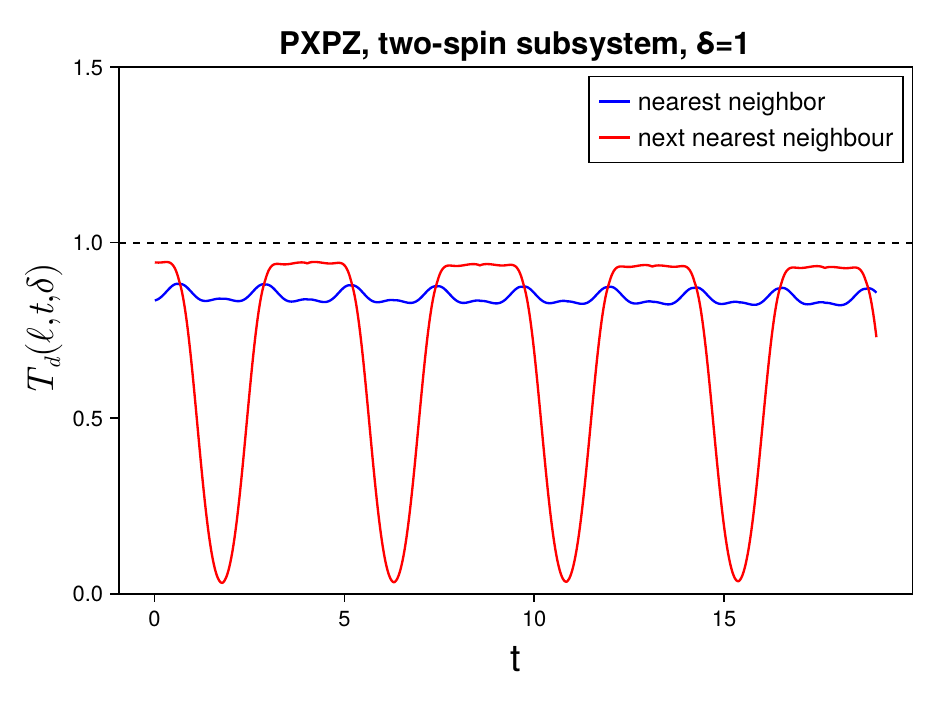}
        \includegraphics[width=5.6cm,height=5cm]{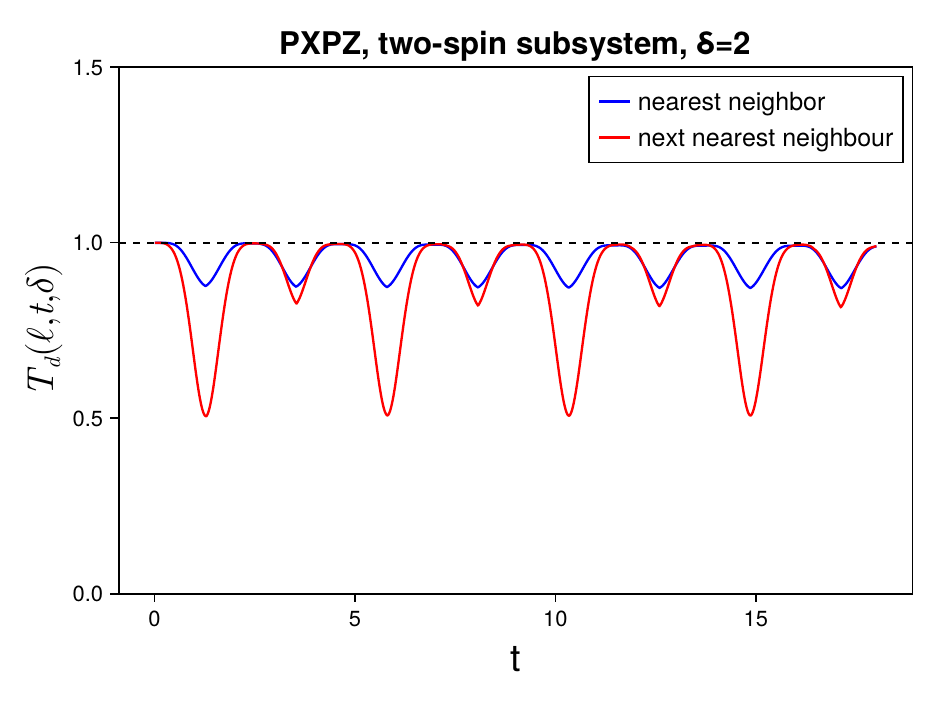}
        \includegraphics[width=5.6cm,height=5cm]{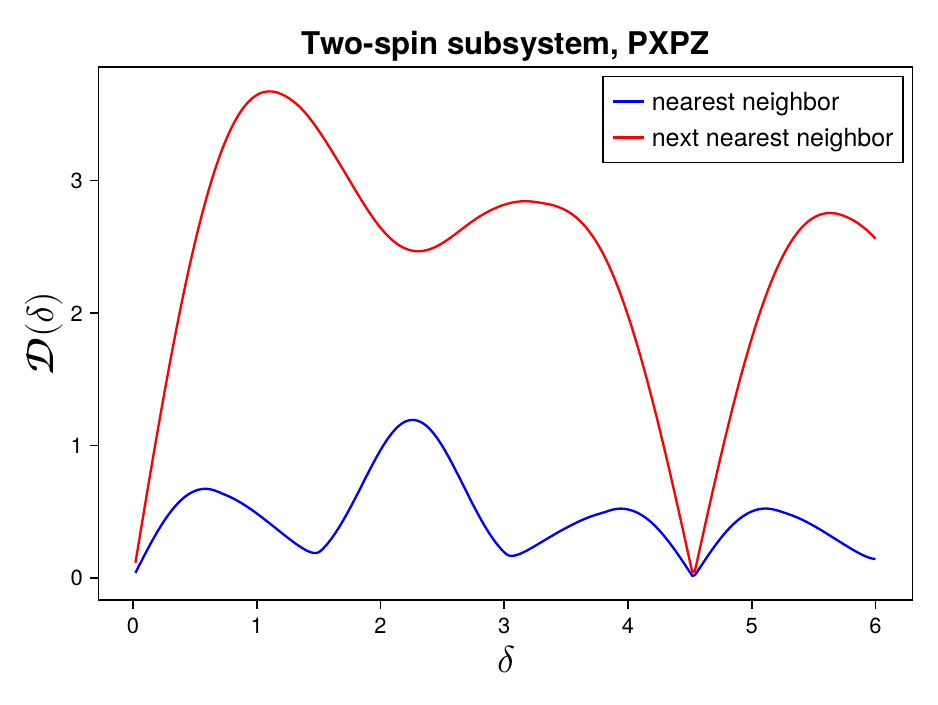}
	\caption{\fontsize{9}{11} \selectfont Comparison of two-spin subsystem's dynamics when the two constituent spins are adjacent/nearest neighbor and next nearest neighbor, specific to the case of the PXPZ model. The dynamics of the latter configuration exhibits significantly higher non-Markovianity.}
	\label{fig:fig7}
\end{figure*}

The actual numerical values of even the maxima of negativity in these cases are quite small. In fact, when the separation considered above is non-zero ($d>0$), the numerical values were seen to be even smaller or even zero for continuous blocks of time at once (even when $d\!=\!1)$.

\section{Weaker non-Markovianity of differently configured subsystems}  \label{appn3}

In the beginning of Sec.\ref{sec4} in the main text, we mentioned that when the constituent spins in a multi-spin subsystem are adjacent (or separated by an even number of sites), the said subsystem exhibits significantly weaker information backflows and non-Markovianity compared to when they are separated by an odd number of sites. Thus, in the main text, we focused on the results with the the latter type of subsystems. In Fig.\ref{fig:fig7}, we demonstrate this contrasting behaviour for the simplest case of a two-spin subsystem comprised of adjacent spins (results are quantitatively unchanged when they are separated by an even number of sites). We choose to show this for only the more interesting case of the PXPZ model in the interests of conciseness.

Similarly, we have seen much weaker non-Markovianity for three- and four-spin subsystems as well when any of their consecutive constituent spins are adjacent or separated by an even number of sites. As mentioned in the main body of this work, these subsystem configurations break the two-site or three-site unit-cell structure of the $\mathbb{Z}_2$ or $\mathbb{Z}_3$ initial states, and the enhanced non-Markovianity may be a direct consequence of this fact. Regrettably, we are unable to provide a satisfactory, even if heuristic, explanation for this observation at this time and hope to properly address this in future.

\section{"Classical" non-monotonicities}    \label{appn4}

\begin{figure*}[!htbp]
	\centering
        \includegraphics[width=5.6cm,height=5cm]{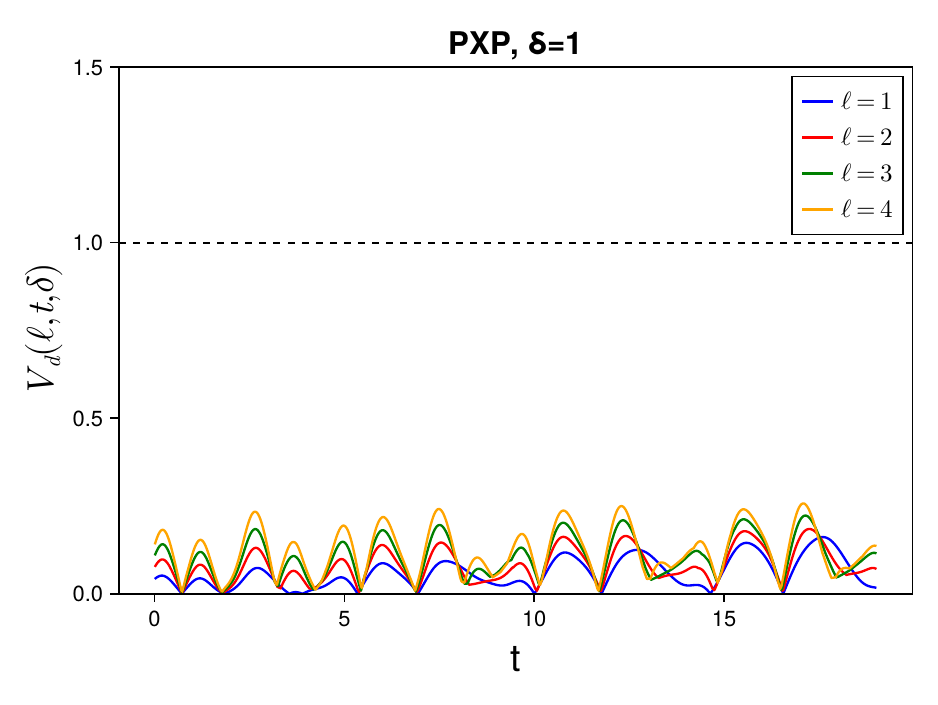}
        \includegraphics[width=5.6cm,height=5cm]{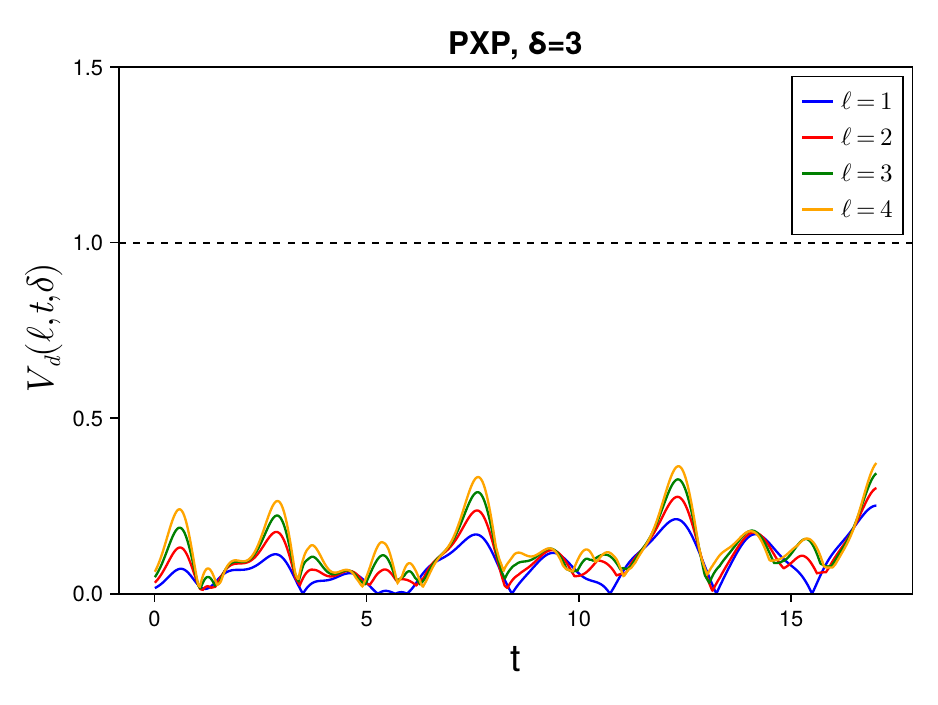}
        \includegraphics[width=5.6cm,height=5cm]{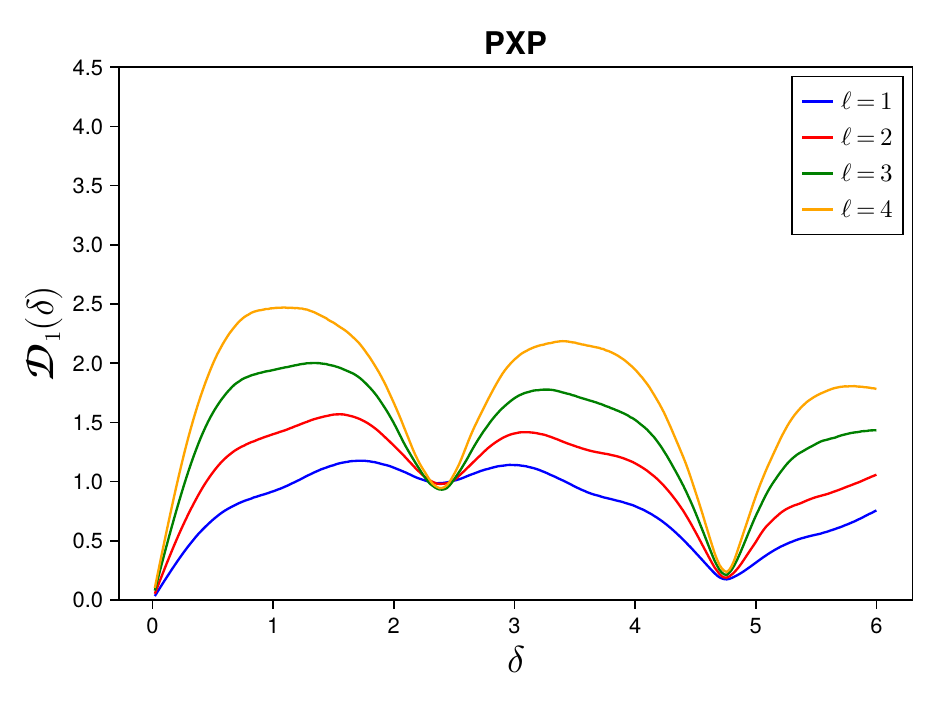}
        \includegraphics[width=5.6cm,height=5cm]{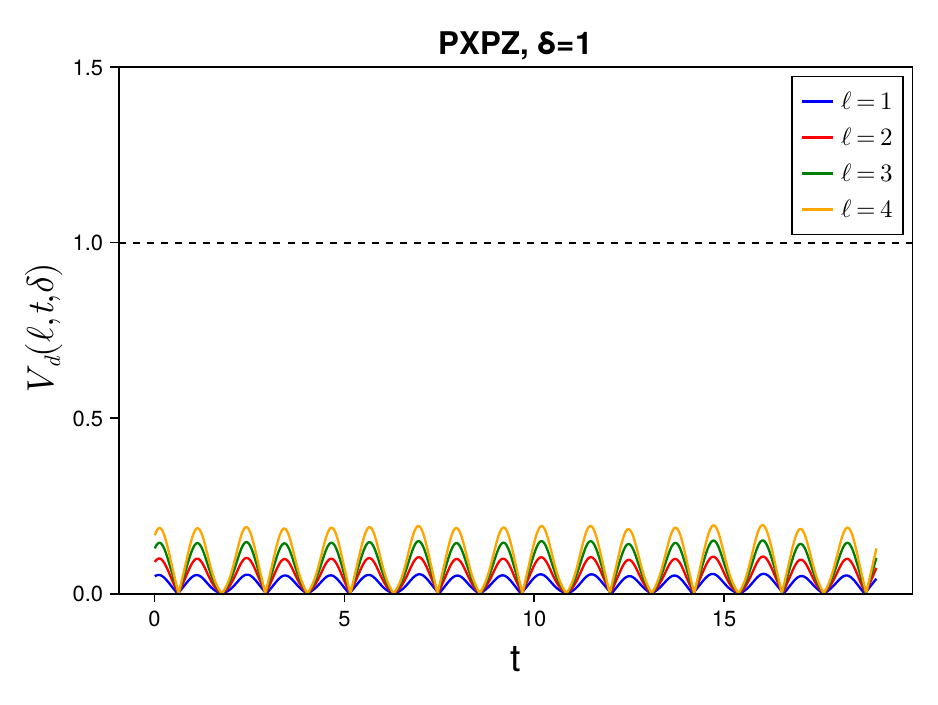}
        \includegraphics[width=5.6cm,height=5cm]{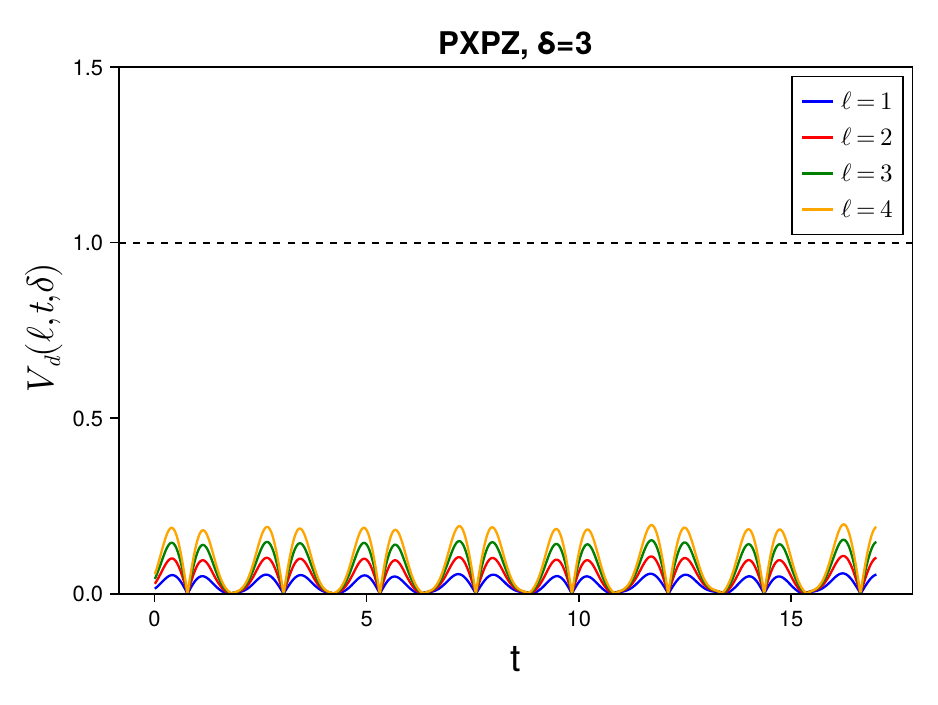}
        \includegraphics[width=5.6cm,height=5cm]{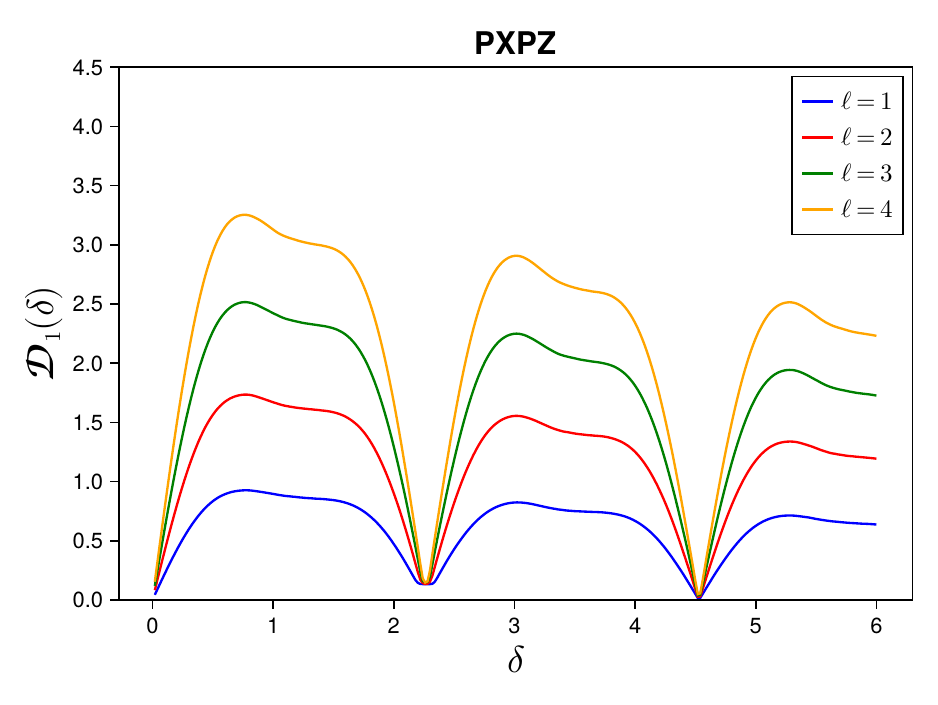}
	\caption{\fontsize{9}{11} \selectfont Behaviour of the TVD, Eq.(\ref{TVD}), and a degree of its revivals, Eq.(\ref{TVDdegree}), for the PXP model (\textbf{upper row}) and the PXPZ model (\textbf{lower row}). A much more systematic behaviour is seen in the PXPZ case. Note the global minima of $\mathcal{D}_1(\delta)$ at $\delta \approx 4.76 $ and $\delta \approx 4.52 $ PXP and PXPZ, respectively. Milder minima at half these values is also apparent. The dependence of $\mathcal{D}_1(\delta)$ on subsystem size is more well-ordered than that of $\mathcal{D}(\delta)$ in Fig.\ref{fig:fig4}.}
	\label{fig:fig8}
\end{figure*}

Many of the metrics measuring distances between quantum states are quantum generalizations of distance metrics between probability distributions. Given any density matrix $\rho$, the set of its eigenvalues $\{p_i\}$ naturally defines a probability distribution. Given any two probability distributions $\{p_i\}$ and $\{q_i\}$, assumed to be ordered in some chosen manner (typically in descending order), one can define a one-norm distance metric called the total variation distance (TVD),
\begin{equation}     \label{TVD}
    V_d(p,q)=\frac{1}{2}\sum_i |p_i - q_i|  \text{   .}
\end{equation}
This can be considered as a "classical" counterpart of the trace distance between two density matrices $\rho$ and $\eta$ whose eigenvalue sets are $\{p_i\}$ and $\{q_i\}$ respectively. Similar to the non-increasing behaviour of trace distances (and other distance metrics between quantum states) under CPTP operations, the TVD (and other distance metrics between probability distributions) is also non-increasing under the so-called data processing operations. Likewise, one may define a degree of TVD revivals as,
\begin{equation}     \label{TVDdegree}
    \mathcal{D}_1(\delta) = \sum_t \alpha_1(t,\delta)  \hspace{0.3cm} \forall t \hspace{0.25cm} \text{s.t.} \hspace{0.25cm} \alpha_1(t,\delta)>0    \text{ ,}
\end{equation}
where,
\begin{equation}    \label{slope1}
    \alpha_1(t,\delta) = \frac{1}{\tau}\bigg(V_d(q_{t+\tau+\delta}^{\ell}, q_{t+\tau}^{\ell}) - V_d(q_{t+\delta}^{\ell}, q_{t}^{\ell}) \bigg)   \text{    ,}
\end{equation}
where $q_{t}^{\ell}$ denotes the eigenvalues, arranged in descending order, of the density matrix $\rho_t^{\ell}$. Hereafter, we denote $V_d(q_{t+\delta}^{\ell}, q_{t}^{\ell})$ with the shorthand $V_d(\ell,t,\delta)$

In Ref.\cite{Banerjee2025a}, we had noted a very systematic behaviour of non-monotonicities of the TVD between descendingly-ordered probability distributions corresponding to the subsystems which exhibited pronounced non-Markovianity as defined in this work in terms of the revivals of trace distances between temporally-separated subsystem states, and had left open the question of whether non-Markovian dynamics of quantum states necessarily also lead to or imply the non-monotonicities (revivals) of TVD, and if so, whether these could be considered a form of "classical memory" as opposed to purely quantum memory of the said dynamical systems. While it is tempting to conjecture this, to the best of our knowledge we are not aware of a firmly established result in the literature that relates non-Markovianity of quantum dynamical systems as measured by information backflows (revivals of quantum distance metrics) to revivals in corresponding classical distance metrics, and subsequently using the latter as a definition or a defining signature of classical memory underlying non-Markovian quantum dynamics. Nonetheless, we believe this is intriguing on its own right and should be investigated separately in future.

In Fig.\ref{fig:fig8}, we show the behaviour of TVD (Eq.(\ref{TVD})) and the degree of TVD revivals (Eq.(\ref{TVDdegree})) for the PXP and PXPZ models. As seen for two example values of $\delta$, the PXPZ model shows a very systematic and clean behaviour of $V_d(\ell,t,\delta)$ (however, their numerical values always remain low). Moreover, almost independently of the subsystem sizes, the degree of TVD revivals $\mathcal{D}_1(\delta)$ has a global minima at $\delta \approx 4.76 $ and $\delta \approx 4.52 $ respectively for the PXP and PXPZ models. Furthermore, unlike for the degree of non-Markovianity $\mathcal{D}(\delta)$ in Fig.\ref{fig:fig4}, there is a cleaner appearance of a milder minima at about half the above values, i.e., $\delta \approx 2.38 $ and $\delta \approx 2.26 $ respectively for the PXP and PXPZ models. In the case of the PXPZ models, even this "milder" minima drops to near zero. In both models, and again unlike $\mathcal{D}(\delta)$ in Fig.\ref{fig:fig4}, a cleanly ordered behaviour is apparent between subsystem sizes, with the larger ones showing more non-monotonic behaviour in $V_d(\ell,t,\delta)$ and subsequently higher $\mathcal{D}_1(\delta)$ than the smaller ones.

\bibliography{biblio}

@article{Turner2018,
  title = {Quantum scarred eigenstates in a Rydberg atom chain: Entanglement, breakdown of thermalization, and stability to perturbations},
  author = {Turner, C. J. and Michailidis, A. A. and Abanin, D. A. and Serbyn, M. and Papi\ifmmode \acute{c}\else \'{c}\fi{}, Z.},
  journal = {Phys. Rev. B},
  volume = {98},
  issue = {15},
  pages = {155134},
  numpages = {23},
  year = {2018},
  month = {Oct},
  publisher = {American Physical Society},
  doi = {10.1103/PhysRevB.98.155134},
  url = {https://link.aps.org/doi/10.1103/PhysRevB.98.155134}
}

@article{Choi2019,
  title = {Emergent SU(2) Dynamics and Perfect Quantum Many-Body Scars},
  author = {Choi, Soonwon and Turner, Christopher J. and Pichler, Hannes and Ho, Wen Wei and Michailidis, Alexios A. and Papi\ifmmode \acute{c}\else \'{c}\fi{}, Zlatko and Serbyn, Maksym and Lukin, Mikhail D. and Abanin, Dmitry A.},
  journal = {Phys. Rev. Lett.},
  volume = {122},
  issue = {22},
  pages = {220603},
  numpages = {6},
  year = {2019},
  month = {Jun},
  publisher = {American Physical Society},
  doi = {10.1103/PhysRevLett.122.220603},
  url = {https://link.aps.org/doi/10.1103/PhysRevLett.122.220603}
}

@article{Khemani2019,
  title = {Signatures of integrability in the dynamics of Rydberg-blockaded chains},
  author = {Khemani, Vedika and Laumann, Chris R. and Chandran, Anushya},
  journal = {Phys. Rev. B},
  volume = {99},
  issue = {16},
  pages = {161101},
  numpages = {6},
  year = {2019},
  month = {Apr},
  publisher = {American Physical Society},
  doi = {10.1103/PhysRevB.99.161101},
  url = {https://link.aps.org/doi/10.1103/PhysRevB.99.161101}
}

@article{Odea2025,
  title = {Entanglement Oscillations from Many-Body Quantum Scars},
  author = {O'Dea, Nicholas and Sriram, Adithya},
  journal = {Phys. Rev. Lett.},
  volume = {134},
  issue = {21},
  pages = {210402},
  numpages = {9},
  year = {2025},
  month = {May},
  publisher = {American Physical Society},
  doi = {10.1103/PhysRevLett.134.210402},
  url = {https://link.aps.org/doi/10.1103/PhysRevLett.134.210402}
}

@article{Dooley2020,
  title = {Enhancing the effect of quantum many-body scars on dynamics by minimizing the effective dimension},
  author = {Dooley, Shane and Kells, Graham},
  journal = {Phys. Rev. B},
  volume = {102},
  issue = {19},
  pages = {195114},
  numpages = {6},
  year = {2020},
  month = {Nov},
  publisher = {American Physical Society},
  doi = {10.1103/PhysRevB.102.195114},
  url = {https://link.aps.org/doi/10.1103/PhysRevB.102.195114}
}

@article{Serbyn2021,
   title={Quantum many-body scars and weak breaking of ergodicity},
   volume={17},
   ISSN={1745-2481},
   url={http://dx.doi.org/10.1038/s41567-021-01230-2},
   DOI={10.1038/s41567-021-01230-2},
   number={6},
   journal={Nature Physics},
   publisher={Springer Science and Business Media LLC},
   author={Serbyn, Maksym and Abanin, Dmitry A. and Papić, Zlatko},
   year={2021},
   month=may, pages={675–685} }

@article{Ho2019,
  title = {Periodic Orbits, Entanglement, and Quantum Many-Body Scars in Constrained Models: Matrix Product State Approach},
  author = {Ho, Wen Wei and Choi, Soonwon and Pichler, Hannes and Lukin, Mikhail D.},
  journal = {Phys. Rev. Lett.},
  volume = {122},
  issue = {4},
  pages = {040603},
  numpages = {6},
  year = {2019},
  month = {Jan},
  publisher = {American Physical Society},
  doi = {10.1103/PhysRevLett.122.040603},
  url = {https://link.aps.org/doi/10.1103/PhysRevLett.122.040603}
}

@article{Alhambra2020,
  title = {Revivals imply quantum many-body scars},
  author = {Alhambra, Alvaro M. and Anshu, Anurag and Wilming, Henrik},
  journal = {Phys. Rev. B},
  volume = {101},
  issue = {20},
  pages = {205107},
  numpages = {12},
  year = {2020},
  month = {May},
  publisher = {American Physical Society},
  doi = {10.1103/PhysRevB.101.205107},
  url = {https://link.aps.org/doi/10.1103/PhysRevB.101.205107}
}

@article{Rozon2024,
  title = {Broken unitary picture of dynamics in quantum many-body scars},
  author = {Rozon, Pierre-Gabriel and Agarwal, Kartiek},
  journal = {Phys. Rev. Res.},
  volume = {6},
  issue = {2},
  pages = {023041},
  numpages = {26},
  year = {2024},
  month = {Apr},
  publisher = {American Physical Society},
  doi = {10.1103/PhysRevResearch.6.023041},
  url = {https://link.aps.org/doi/10.1103/PhysRevResearch.6.023041}
}

@book{Wilde2017, place={Cambridge}, title={Quantum Information Theory}, publisher={Cambridge University Press}, author={Wilde, Mark M.}, year={2017}}

@book{Hayashi2017,
  title={Quantum Information Theory: Mathematical Foundation},
  author={Hayashi, M.},
  isbn={9783662435014},
  year={2017},
  publisher={Springer-Verlag}
}

@book{Alicki2007,
  title={Quantum dynamical semigroups and applications},
  author={Alicki, Robert and Lendi, Karl},
  volume={717},
  year={2007},
  publisher={Springer}
}

@article{Chruscinski2022,
title = {Dynamical maps beyond Markovian regime},
journal = {Physics Reports},
volume = {992},
pages = {1-85},
year = {2022},
issn = {0370-1573},
doi = {https://doi.org/10.1016/j.physrep.2022.09.003},
url = {https://www.sciencedirect.com/science/article/pii/S0370157322003428},
author = {Dariusz Chruściński}
}

@article{Breuer2009,
  title = {Measure for the Degree of Non-Markovian Behavior of Quantum Processes in Open Systems},
  author = {Breuer, Heinz-Peter and Laine, Elsi-Mari and Piilo, Jyrki},
  journal = {Phys. Rev. Lett.},
  volume = {103},
  issue = {21},
  pages = {210401},
  numpages = {4},
  year = {2009},
  month = {Nov},
  publisher = {American Physical Society},
  doi = {10.1103/PhysRevLett.103.210401},
  url = {https://link.aps.org/doi/10.1103/PhysRevLett.103.210401}
}

@article{Breuer2010,
  title = {Measure for the non-Markovianity of quantum processes},
  author = {Laine, Elsi-Mari and Piilo, Jyrki and Breuer, Heinz-Peter},
  journal = {Phys. Rev. A},
  volume = {81},
  issue = {6},
  pages = {062115},
  numpages = {8},
  year = {2010},
  month = {Jun},
  publisher = {American Physical Society},
  doi = {10.1103/PhysRevA.81.062115},
  url = {https://link.aps.org/doi/10.1103/PhysRevA.81.062115}
}

@article{Bernien2017,
   title={Probing many-body dynamics on a 51-atom quantum simulator},
   volume={551},
   ISSN={1476-4687},
   url={http://dx.doi.org/10.1038/nature24622},
   DOI={10.1038/nature24622},
   number={7682},
   journal={Nature},
   publisher={Springer Science and Business Media LLC},
   author={Bernien, Hannes and Schwartz, Sylvain and Keesling, Alexander and Levine, Harry and Omran, Ahmed and Pichler, Hannes and Choi, Soonwon and Zibrov, Alexander S. and Endres, Manuel and Greiner, Markus and Vuletić, Vladan and Lukin, Mikhail D.},
   year={2017},
   month=nov, pages={579–584} 
}

@article{Turner2017,
   title={Weak ergodicity breaking from quantum many-body scars},
   volume={14},
   ISSN={1745-2481},
   url={http://dx.doi.org/10.1038/s41567-018-0137-5},
   DOI={10.1038/s41567-018-0137-5},
   number={7},
   journal={Nature Physics},
   publisher={Springer Science and Business Media LLC},
   author={Turner, C. J. and Michailidis, A. A. and Abanin, D. A. and Serbyn, M. and Papić, Z.},
   year={2018},
   month=may, pages={745–749} 
}

@article{Lin2019,
   title={Exact Quantum Many-Body Scar States in the Rydberg-Blockaded Atom Chain},
   volume={122},
   ISSN={1079-7114},
   url={http://dx.doi.org/10.1103/PhysRevLett.122.173401},
   DOI={10.1103/physrevlett.122.173401},
   number={17},
   journal={Physical Review Letters},
   publisher={American Physical Society (APS)},
   author={Lin, Cheng-Ju and Motrunich, Olexei I.},
   year={2019},
   month=apr 
}

@article{Banerjee2025a,
  title = {Non-Markovianity of subsystem dynamics in isolated quantum many-body systems},
  author = {Banerjee, Aditya},
  journal = {Phys. Rev. B},
  volume = {112},
  issue = {1},
  pages = {014302},
  numpages = {10},
  year = {2025},
  month = {Jul},
  publisher = {American Physical Society},
  doi = {10.1103/q8h3-mkzr},
  url = {https://link.aps.org/doi/10.1103/q8h3-mkzr}
}

@article{Moudgalya2022,
doi = {10.1088/1361-6633/ac73a0},
url = {https://dx.doi.org/10.1088/1361-6633/ac73a0},
year = {2022},
month = {jul},
publisher = {IOP Publishing},
volume = {85},
number = {8},
pages = {086501},
author = {Sanjay Moudgalya and B Andrei Bernevig and Nicolas Regnault},
title = {Quantum many-body scars and Hilbert space fragmentation: a review of exact results},
journal = {Reports on Progress in Physics}
}

@misc{Ivanov2025,
      title={Many exact area-law scar eigenstates in the nonintegrable PXP and related models}, 
      author={Andrew N. Ivanov and Olexei I. Motrunich},
      year={2025},
      eprint={2503.16327},
      archivePrefix={arXiv},
      primaryClass={quant-ph},
      url={https://arxiv.org/abs/2503.16327}, 
}

@article{Ivanov2024,
  title = {Volume-Entangled Exact Scar States in the PXP and Related Models in Any Dimension},
  author = {Ivanov, Andrew N. and Motrunich, Olexei I.},
  journal = {Phys. Rev. Lett.},
  volume = {134},
  issue = {5},
  pages = {050403},
  numpages = {6},
  year = {2025},
  month = {Feb},
  publisher = {American Physical Society},
  doi = {10.1103/PhysRevLett.134.050403},
  url = {https://link.aps.org/doi/10.1103/PhysRevLett.134.050403}
}

@article{Ljubotina2023,
  title = {Superdiffusive Energy Transport in Kinetically Constrained Models},
  author = {Ljubotina, Marko and Desaules, Jean-Yves and Serbyn, Maksym and Papic, Zlatko},
  journal = {Phys. Rev. X},
  volume = {13},
  issue = {1},
  pages = {011033},
  numpages = {14},
  year = {2023},
  month = {Mar},
  publisher = {American Physical Society},
  doi = {10.1103/PhysRevX.13.011033},
  url = {https://link.aps.org/doi/10.1103/PhysRevX.13.011033}
}

@article{Romero2019,
   title={Almost Markovian processes from closed dynamics},
   volume={3},
   ISSN={2521-327X},
   url={http://dx.doi.org/10.22331/q-2019-04-30-136},
   DOI={10.22331/q-2019-04-30-136},
   journal={Quantum},
   publisher={Verein zur Forderung des Open Access Publizierens in den Quantenwissenschaften},
   author={Figueroa-Romero, Pedro and Modi, Kavan and Pollock, Felix A.},
   year={2019},
   month=apr, pages={136} 
}

@article{Omiya2023,
   title={Quantum many-body scars in bipartite Rydberg arrays originating from hidden projector embedding},
   volume={107},
   ISSN={2469-9934},
   url={http://dx.doi.org/10.1103/PhysRevA.107.023318},
   DOI={10.1103/physreva.107.023318},
   number={2},
   journal={Physical Review A},
   publisher={American Physical Society (APS)},
   author={Omiya, Keita and Müller, Markus},
   year={2023},
   month=feb 
}

@article{Bull2020,
  title = {Quantum scars as embeddings of weakly broken Lie algebra representations},
  author = {Bull, Kieran and Desaules, Jean-Yves and Papic, Zlatko},
  journal = {Phys. Rev. B},
  volume = {101},
  issue = {16},
  pages = {165139},
  numpages = {17},
  year = {2020},
  month = {Apr},
  publisher = {American Physical Society},
  doi = {10.1103/PhysRevB.101.165139},
  url = {https://link.aps.org/doi/10.1103/PhysRevB.101.165139}
}

@article{Kerschbaumer2025,
  title = {Quantum Many-Body Scars beyond the PXP Model in Rydberg Simulators},
  author = {Kerschbaumer, Aron and Ljubotina, Marko and Serbyn, Maksym and Desaules, Jean-Yves},
  journal = {Phys. Rev. Lett.},
  volume = {134},
  issue = {16},
  pages = {160401},
  numpages = {9},
  year = {2025},
  month = {Apr},
  publisher = {American Physical Society},
  doi = {10.1103/PhysRevLett.134.160401},
  url = {https://link.aps.org/doi/10.1103/PhysRevLett.134.160401}
}

@article{Schnee2025,
   title={Unconventional early-time relaxation in the Rydberg chain},
   volume={113},
   ISSN={2469-9969},
   url={http://dx.doi.org/10.1103/q3q8-gssw},
   DOI={10.1103/q3q8-gssw},
   number={8},
   journal={Physical Review B},
   publisher={American Physical Society (APS)},
   author={Schnee, Martin and Radgohar, Roya and Kourtis, Stefanos},
   year={2026},
   month=feb }

@article{Gross2021,
   title={Quantum gas microscopy for single atom and spin detection},
   volume={17},
   ISSN={1745-2481},
   url={http://dx.doi.org/10.1038/s41567-021-01370-5},
   DOI={10.1038/s41567-021-01370-5},
   number={12},
   journal={Nature Physics},
   publisher={Springer Science and Business Media LLC},
   author={Gross, Christian and Bakr, Waseem S.},
   year={2021},
   month=nov, pages={1316–1323} 
}

@article{Kuhr2016,
   title={Quantum-gas microscopes: a new tool for cold-atom quantum simulators},
   volume={3},
   ISSN={2095-5138},
   url={http://dx.doi.org/10.1093/nsr/nww023},
   DOI={10.1093/nsr/nww023},
   number={2},
   journal={National Science Review},
   publisher={Oxford University Press (OUP)},
   author={Kuhr, Stefan},
   year={2016},
   month=apr, pages={170–172} 
}

@article{Ott2016,
   title={Single atom detection in ultracold quantum gases: a review of current progress},
   volume={79},
   ISSN={1361-6633},
   url={http://dx.doi.org/10.1088/0034-4885/79/5/054401},
   DOI={10.1088/0034-4885/79/5/054401},
   number={5},
   journal={Reports on Progress in Physics},
   publisher={IOP Publishing},
   author={Ott, Herwig},
   year={2016},
   month=apr, pages={054401} 
}

@article{ITensor,
	title={{The ITensor Software Library for Tensor Network Calculations}},
	author={Matthew Fishman and Steven R. White and E. Miles Stoudenmire},
	journal={SciPost Phys. Codebases},
	pages={4},
	year={2022},
	publisher={SciPost},
	doi={10.21468/SciPostPhysCodeb.4},
	url={https://scipost.org/10.21468/SciPostPhysCodeb.4},
}

@article{Vidal2004,
  title = {Efficient Simulation of One-Dimensional Quantum Many-Body Systems},
  author = {Vidal, Guifr\'e},
  journal = {Phys. Rev. Lett.},
  volume = {93},
  issue = {4},
  pages = {040502},
  numpages = {4},
  year = {2004},
  month = {Jul},
  publisher = {American Physical Society},
  doi = {10.1103/PhysRevLett.93.040502},
  url = {https://link.aps.org/doi/10.1103/PhysRevLett.93.040502}
}

@article{Giarmatzi2021,
   title={Witnessing quantum memory in non-Markovian processes},
   volume={5},
   ISSN={2521-327X},
   url={http://dx.doi.org/10.22331/q-2021-04-26-440},
   DOI={10.22331/q-2021-04-26-440},
   journal={Quantum},
   publisher={Verein zur Forderung des Open Access Publizierens in den Quantenwissenschaften},
   author={Giarmatzi, Christina and Costa, Fabio},
   year={2021},
   month=apr, pages={440} 
}

@article{Milz2020,
  title = {When Is a Non-Markovian Quantum Process Classical?},
  author = {Milz, Simon and Egloff, Dario and Taranto, Philip and Theurer, Thomas and Plenio, Martin B. and Smirne, Andrea and Huelga, Susana F.},
  journal = {Phys. Rev. X},
  volume = {10},
  issue = {4},
  pages = {041049},
  numpages = {42},
  year = {2020},
  month = {Dec},
  publisher = {American Physical Society},
  doi = {10.1103/PhysRevX.10.041049},
  url = {https://link.aps.org/doi/10.1103/PhysRevX.10.041049}
}

@article{Backer2024,
  title = {Local Disclosure of Quantum Memory in Non-Markovian Dynamics},
  author = {B\"acker, Charlotte and Beyer, Konstantin and Strunz, Walter T.},
  journal = {Phys. Rev. Lett.},
  volume = {132},
  issue = {6},
  pages = {060402},
  numpages = {6},
  year = {2024},
  month = {Feb},
  publisher = {American Physical Society},
  doi = {10.1103/PhysRevLett.132.060402},
  url = {https://link.aps.org/doi/10.1103/PhysRevLett.132.060402}
}

@article{Backer2025,
  title = {Entropic witness for quantum memory in open system dynamics},
  author = {B\"acker, Charlotte and Beyer, Konstantin and Strunz, Walter T.},
  journal = {Phys. Rev. Res.},
  volume = {7},
  issue = {3},
  pages = {033256},
  numpages = {10},
  year = {2025},
  month = {Sep},
  publisher = {American Physical Society},
  doi = {10.1103/618n-fp8w},
  url = {https://link.aps.org/doi/10.1103/618n-fp8w}
}

@article{Bravo2025,
   title={Purely quantum memory in closed systems observed via imperfect measurements},
   volume={9},
   ISSN={2521-327X},
   url={http://dx.doi.org/10.22331/q-2025-12-11-1938},
   DOI={10.22331/q-2025-12-11-1938},
   journal={Quantum},
   publisher={Verein zur Forderung des Open Access Publizierens in den Quantenwissenschaften},
   author={Tabanera-Bravo, Jorge and Godec, Aljaz},
   year={2025},
   month=dec, pages={1938} }

@article{Banacki2023,
  title = {Information backflow may not indicate quantum memory},
  author = {Banacki, Micha\l{} and Marciniak, Marcin and Horodecki, Karol and Horodecki, Pawe\l{}},
  journal = {Phys. Rev. A},
  volume = {107},
  issue = {3},
  pages = {032202},
  numpages = {7},
  year = {2023},
  month = {Mar},
  publisher = {American Physical Society},
  doi = {10.1103/PhysRevA.107.032202},
  url = {https://link.aps.org/doi/10.1103/PhysRevA.107.032202}
}

@article{Buscemi2025,
   title={Causal and Noncausal Revivals of Information: A New Regime of Non-Markovianity in Quantum Stochastic Processes},
   volume={6},
   ISSN={2691-3399},
   url={http://dx.doi.org/10.1103/PRXQuantum.6.020316},
   DOI={10.1103/prxquantum.6.020316},
   number={2},
   journal={PRX Quantum},
   publisher={American Physical Society (APS)},
   author={Buscemi, Francesco and Gangwar, Rajeev and Goswami, Kaumudibikash and Badhani, Himanshu and Pandit, Tanmoy and Mohan, Brij and Das, Siddhartha and Bera, Manabendra Nath},
   year={2025},
   month=apr }

@article{Cerrillo2014,
  title = {Non-Markovian Dynamical Maps: Numerical Processing of Open Quantum Trajectories},
  author = {Cerrillo, Javier and Cao, Jianshu},
  journal = {Phys. Rev. Lett.},
  volume = {112},
  issue = {11},
  pages = {110401},
  numpages = {5},
  year = {2014},
  month = {Mar},
  publisher = {American Physical Society},
  doi = {10.1103/PhysRevLett.112.110401},
  url = {https://link.aps.org/doi/10.1103/PhysRevLett.112.110401}
}

@article{Rosenbach2016,
   title={Efficient simulation of non-Markovian system-environment interaction},
   volume={18},
   ISSN={1367-2630},
   url={http://dx.doi.org/10.1088/1367-2630/18/2/023035},
   DOI={10.1088/1367-2630/18/2/023035},
   number={2},
   journal={New Journal of Physics},
   publisher={IOP Publishing},
   author={Rosenbach, Robert and Cerrillo, Javier and Huelga, Susana F and Cao, Jianshu and Plenio, Martin B},
   year={2016},
   month=feb, pages={023035} 
}

@article{Pollock2018,
   title={Tomographically reconstructed master equations for any open quantum dynamics},
   volume={2},
   ISSN={2521-327X},
   url={http://dx.doi.org/10.22331/q-2018-07-11-76},
   DOI={10.22331/q-2018-07-11-76},
   journal={Quantum},
   publisher={Verein zur Forderung des Open Access Publizierens in den Quantenwissenschaften},
   author={Pollock, Felix A. and Modi, Kavan},
   year={2018},
   month=jul, pages={76} 
}

@article{Strachan2024,
   title={Extracting dynamical maps of non-Markovian open quantum systems},
   volume={161},
   ISSN={1089-7690},
   url={http://dx.doi.org/10.1063/5.0228428},
   DOI={10.1063/5.0228428},
   number={15},
   journal={The Journal of Chemical Physics},
   publisher={AIP Publishing},
   author={Strachan, David J. and Purkayastha, Archak and Clark, Stephen R.},
   year={2024},
   month=oct 
}

@article{Coppola2025,
   title={Learning the non-Markovian features of subsystem dynamics},
   volume={19},
   ISSN={2542-4653},
   url={http://dx.doi.org/10.21468/SciPostPhys.19.6.149},
   DOI={10.21468/scipostphys.19.6.149},
   number={6},
   journal={SciPost Physics},
   publisher={Stichting SciPost},
   author={Coppola, Michele and Bañuls, Mari Carmen and Lenarčič, Zala},
   year={2025},
   month=dec }

@article{Gogolin2015,
   title={Equilibration, thermalisation, and the emergence of statistical mechanics in closed quantum systems},
   volume={79},
   ISSN={1361-6633},
   url={http://dx.doi.org/10.1088/0034-4885/79/5/056001},
   DOI={10.1088/0034-4885/79/5/056001},
   number={5},
   journal={Reports on Progress in Physics},
   publisher={IOP Publishing},
   author={Gogolin, Christian and Eisert, Jens},
   year={2016},
   month=apr, pages={056001} }

@article{Deutsch2018,
   title={Eigenstate thermalization hypothesis},
   volume={81},
   ISSN={1361-6633},
   url={http://dx.doi.org/10.1088/1361-6633/aac9f1},
   DOI={10.1088/1361-6633/aac9f1},
   number={8},
   journal={Reports on Progress in Physics},
   publisher={IOP Publishing},
   author={Deutsch, Joshua M},
   year={2018},
   month=jul, pages={082001} }

@article{Chandran2023,
   author = "Chandran, Anushya and Iadecola, Thomas and Khemani, Vedika and Moessner, Roderich",
   title = "Quantum Many-Body Scars: A Quasiparticle Perspective", 
   journal= "Annual Review of Condensed Matter Physics",
   year = "2023",
   volume = "14",
   number = "Volume 14, 2023",
   pages = "443-469",
   doi = "https://doi.org/10.1146/annurev-conmatphys-031620-101617",
   url = "https://www.annualreviews.org/content/journals/10.1146/annurev-conmatphys-031620-101617",
   publisher = "Annual Reviews",
   issn = "1947-5462",
   type = "Journal Article"
  }

@book{Vacchini2024,
  title={Open Quantum Systems: Foundations and Theory},
  author={Vacchini, Bassano},
  year={2024},
  publisher={Springer Nature}
}

@article{Rivas2014,
   title={Quantum non-Markovianity: characterization, quantification and detection},
   volume={77},
   ISSN={1361-6633},
   url={http://dx.doi.org/10.1088/0034-4885/77/9/094001},
   DOI={10.1088/0034-4885/77/9/094001},
   number={9},
   journal={Reports on Progress in Physics},
   publisher={IOP Publishing},
   author={Rivas, Angel and Huelga, Susana F and Plenio, Martin B},
   year={2014},
   month=aug, pages={094001} 
}

@article{Breuer2016,
   title={Colloquium: Non-Markovian dynamics in open quantum systems},
   volume={88},
   ISSN={1539-0756},
   url={http://dx.doi.org/10.1103/RevModPhys.88.021002},
   DOI={10.1103/revmodphys.88.021002},
   number={2},
   journal={Reviews of Modern Physics},
   publisher={American Physical Society (APS)},
   author={Breuer, Heinz-Peter and Laine, Elsi-Mari and Piilo, Jyrki and Vacchini, Bassano},
   year={2016},
   month=apr 
}

@article{Li2018,
   title={Concepts of quantum non-Markovianity: A hierarchy},
   volume={759},
   ISSN={0370-1573},
   url={http://dx.doi.org/10.1016/j.physrep.2018.07.001},
   DOI={10.1016/j.physrep.2018.07.001},
   journal={Physics Reports},
   publisher={Elsevier BV},
   author={Li, Li and Hall, Michael J.W. and Wiseman, Howard M.},
   year={2018},
   month=oct, 
   pages={1–51} 
}

@article{Vega2017,
  title = {Dynamics of non-Markovian open quantum systems},
  author = {de Vega, In\'es and Alonso, Daniel},
  journal = {Rev. Mod. Phys.},
  volume = {89},
  issue = {1},
  pages = {015001},
  numpages = {58},
  year = {2017},
  month = {Jan},
  publisher = {American Physical Society},
  doi = {10.1103/RevModPhys.89.015001},
  url = {https://link.aps.org/doi/10.1103/RevModPhys.89.015001}
}

@article{Kormos2016,
   title={Real-time confinement following a quantum quench to a non-integrable model},
   volume={13},
   ISSN={1745-2481},
   url={http://dx.doi.org/10.1038/nphys3934},
   DOI={10.1038/nphys3934},
   number={3},
   journal={Nature Physics},
   publisher={Springer Science and Business Media LLC},
   author={Kormos, Marton and Collura, Mario and Takács, Gabor and Calabrese, Pasquale},
   year={2016},
   month=nov, pages={246–249} }

@article{Birnkammer2022,
   title={Prethermalization in one-dimensional quantum many-body systems with confinement},
   volume={13},
   ISSN={2041-1723},
   url={http://dx.doi.org/10.1038/s41467-022-35301-6},
   DOI={10.1038/s41467-022-35301-6},
   number={1},
   journal={Nature Communications},
   publisher={Springer Science and Business Media LLC},
   author={Birnkammer, Stefan and Bastianello, Alvise and Knap, Michael},
   year={2022},
   month=dec }

@article{James2019,
  title = {Nonthermal States Arising from Confinement in One and Two Dimensions},
  author = {James, Andrew J. A. and Konik, Robert M. and Robinson, Neil J.},
  journal = {Phys. Rev. Lett.},
  volume = {122},
  issue = {13},
  pages = {130603},
  numpages = {8},
  year = {2019},
  month = {Apr},
  publisher = {American Physical Society},
  doi = {10.1103/PhysRevLett.122.130603},
  url = {https://link.aps.org/doi/10.1103/PhysRevLett.122.130603}
}

@article{Peres1996,
  title = {Separability Criterion for Density Matrices},
  author = {Peres, Asher},
  journal = {Phys. Rev. Lett.},
  volume = {77},
  issue = {8},
  pages = {1413--1415},
  numpages = {0},
  year = {1996},
  month = {Aug},
  publisher = {American Physical Society},
  doi = {10.1103/PhysRevLett.77.1413},
  url = {https://link.aps.org/doi/10.1103/PhysRevLett.77.1413}
}

@article{Horodeckis1996,
   title={Separability of mixed states: necessary and sufficient conditions},
   volume={223},
   ISSN={0375-9601},
   url={http://dx.doi.org/10.1016/S0375-9601(96)00706-2},
   DOI={10.1016/s0375-9601(96)00706-2},
   number={1–2},
   journal={Physics Letters A},
   publisher={Elsevier BV},
   author={Horodecki, Michał and Horodecki, Paweł and Horodecki, Ryszard},
   year={1996},
   month=nov, pages={1–8} }

@article{VidalWerner2002,
  title = {Computable measure of entanglement},
  author = {Vidal, G. and Werner, R. F.},
  journal = {Phys. Rev. A},
  volume = {65},
  issue = {3},
  pages = {032314},
  numpages = {11},
  year = {2002},
  month = {Feb},
  publisher = {American Physical Society},
  doi = {10.1103/PhysRevA.65.032314},
  url = {https://link.aps.org/doi/10.1103/PhysRevA.65.032314}
}

@article{Pollock2018a,
  title = {Non-Markovian quantum processes: Complete framework and efficient characterization},
  author = {Pollock, Felix A. and Rodr\'{\i}guez-Rosario, C\'esar and Frauenheim, Thomas and Paternostro, Mauro and Modi, Kavan},
  journal = {Phys. Rev. A},
  volume = {97},
  issue = {1},
  pages = {012127},
  numpages = {13},
  year = {2018},
  month = {Jan},
  publisher = {American Physical Society},
  doi = {10.1103/PhysRevA.97.012127},
  url = {https://link.aps.org/doi/10.1103/PhysRevA.97.012127}
}

@article{Guo2020,
  title = {Tensor-network-based machine learning of non-Markovian quantum processes},
  author = {Guo, Chu and Modi, Kavan and Poletti, Dario},
  journal = {Phys. Rev. A},
  volume = {102},
  issue = {6},
  pages = {062414},
  numpages = {8},
  year = {2020},
  month = {Dec},
  publisher = {American Physical Society},
  doi = {10.1103/PhysRevA.102.062414},
  url = {https://link.aps.org/doi/10.1103/PhysRevA.102.062414}
}

@article{Guo2022,
   title={Quantifying Non-Markovianity in Open Quantum Dynamics},
   volume={13},
   ISSN={2542-4653},
   url={http://dx.doi.org/10.21468/SciPostPhys.13.2.028},
   DOI={10.21468/scipostphys.13.2.028},
   number={2},
   journal={SciPost Physics},
   publisher={Stichting SciPost},
   author={Guo, Chu},
   year={2022},
   month=aug }

@article{Ciccarello2022,
   title={Quantum collision models: Open system dynamics from repeated interactions},
   volume={954},
   ISSN={0370-1573},
   url={http://dx.doi.org/10.1016/j.physrep.2022.01.001},
   DOI={10.1016/j.physrep.2022.01.001},
   journal={Physics Reports},
   publisher={Elsevier BV},
   author={Ciccarello, Francesco and Lorenzo, Salvatore and Giovannetti, Vittorio and Palma, G. Massimo},
   year={2022},
   month=apr, pages={1–70} }

@article{Michailidis2020,
  title = {Slow Quantum Thermalization and Many-Body Revivals from Mixed Phase Space},
  author = {Michailidis, A. A. and Turner, C. J. and Papi\ifmmode \acute{c}\else \'{c}\fi{}, Z. and Abanin, D. A. and Serbyn, M.},
  journal = {Phys. Rev. X},
  volume = {10},
  issue = {1},
  pages = {011055},
  numpages = {28},
  year = {2020},
  month = {Mar},
  publisher = {American Physical Society},
  doi = {10.1103/PhysRevX.10.011055},
  url = {https://link.aps.org/doi/10.1103/PhysRevX.10.011055}
}

@article{Pizzi2024,
   title={Genuine quantum scars in many-body spin systems},
   volume={16},
   ISSN={2041-1723},
   url={http://dx.doi.org/10.1038/s41467-025-61765-3},
   DOI={10.1038/s41467-025-61765-3},
   number={1},
   journal={Nature Communications},
   publisher={Springer Science and Business Media LLC},
   author={Pizzi, Andrea and Kwan, Long-Hei and Evrard, Bertrand and Dag, Ceren B. and Knolle, Johannes},
   year={2025},
   month=jul }

@article{Pizzi2025,
  title = {Quantum Trails and Memory Effects in the Phase Space of Chaotic Quantum Systems},
  author = {Pizzi, Andrea},
  journal = {Phys. Rev. Lett.},
  volume = {134},
  issue = {14},
  pages = {140402},
  numpages = {7},
  year = {2025},
  month = {Apr},
  publisher = {American Physical Society},
  doi = {10.1103/PhysRevLett.134.140402},
  url = {https://link.aps.org/doi/10.1103/PhysRevLett.134.140402}
}

@misc{Graf2024,
      title={Birthmarks: Ergodicity Breaking Beyond Quantum Scars}, 
      author={Anton M. Graf and Joonas Keski-Rahkonen and Mingxuan Xiao and Saul Atwood and Zhongling Lu and Siyuan Chen and Eric J. Heller},
      year={2024},
      eprint={2412.02982},
      archivePrefix={arXiv},
      primaryClass={quant-ph},
      url={https://arxiv.org/abs/2412.02982}, 
}

@article{Evrard2024,
  title = {Quantum many-body scars from unstable periodic orbits},
  author = {Evrard, Bertrand and Pizzi, Andrea and Mistakidis, Simeon I. and Dag, Ceren B.},
  journal = {Phys. Rev. B},
  volume = {110},
  issue = {14},
  pages = {144302},
  numpages = {15},
  year = {2024},
  month = {Oct},
  publisher = {American Physical Society},
  doi = {10.1103/PhysRevB.110.144302},
  url = {https://link.aps.org/doi/10.1103/PhysRevB.110.144302}
}

@article{Ermakov2024,
   title={Classical periodic trajectories and quantum scars in many-spin systems},
   volume={112},
   ISSN={2470-0053},
   url={http://dx.doi.org/10.1103/zcgw-q34x},
   DOI={10.1103/zcgw-q34x},
   number={6},
   journal={Physical Review E},
   publisher={American Physical Society (APS)},
   author={Ermakov, Igor and Lychkovskiy, Oleg and Fine, Boris V.},
   year={2025},
   month=dec }

@article{Shen2024,
  title = {Enhanced Many-Body Quantum Scars from the Non-Hermitian Fock Skin Effect},
  author = {Shen, Ruizhe and Qin, Fang and Desaules, Jean-Yves and Papi\ifmmode \acute{c}\else \'{c}\fi{}, Zlatko and Lee, Ching Hua},
  journal = {Phys. Rev. Lett.},
  volume = {133},
  issue = {21},
  pages = {216601},
  numpages = {9},
  year = {2024},
  month = {Nov},
  publisher = {American Physical Society},
  doi = {10.1103/PhysRevLett.133.216601},
  url = {https://link.aps.org/doi/10.1103/PhysRevLett.133.216601}
}

@article{Chen2023,
   title={Weak ergodicity breaking in non-Hermitian many-body systems},
   volume={15},
   ISSN={2542-4653},
   url={http://dx.doi.org/10.21468/SciPostPhys.15.2.052},
   DOI={10.21468/scipostphys.15.2.052},
   number={2},
   journal={SciPost Physics},
   publisher={Stichting SciPost},
   author={Chen, Qianqian and Chen, Shuai A. and Zhu, Zheng},
   year={2023},
   month=aug }

@article{Reimann2008,
  title = {Foundation of Statistical Mechanics under Experimentally Realistic Conditions},
  author = {Reimann, Peter},
  journal = {Phys. Rev. Lett.},
  volume = {101},
  issue = {19},
  pages = {190403},
  numpages = {4},
  year = {2008},
  month = {Nov},
  publisher = {American Physical Society},
  doi = {10.1103/PhysRevLett.101.190403},
  url = {https://link.aps.org/doi/10.1103/PhysRevLett.101.190403}
}

@article{Linden2009,
  title = {Quantum mechanical evolution towards thermal equilibrium},
  author = {Linden, Noah and Popescu, Sandu and Short, Anthony J. and Winter, Andreas},
  journal = {Phys. Rev. E},
  volume = {79},
  issue = {6},
  pages = {061103},
  numpages = {12},
  year = {2009},
  month = {Jun},
  publisher = {American Physical Society},
  doi = {10.1103/PhysRevE.79.061103},
  url = {https://link.aps.org/doi/10.1103/PhysRevE.79.061103}
}

@article{Short2010,
   title={Equilibration of quantum systems and subsystems},
   volume={13},
   ISSN={1367-2630},
   url={http://dx.doi.org/10.1088/1367-2630/13/5/053009},
   DOI={10.1088/1367-2630/13/5/053009},
   number={5},
   journal={New Journal of Physics},
   publisher={IOP Publishing},
   author={Short, Anthony J},
   year={2011},
   month=may, pages={053009} 
}

@article{Short2011,
   title={Quantum equilibration in finite time},
   volume={14},
   ISSN={1367-2630},
   url={http://dx.doi.org/10.1088/1367-2630/14/1/013063},
   DOI={10.1088/1367-2630/14/1/013063},
   number={1},
   journal={New Journal of Physics},
   publisher={IOP Publishing},
   author={Short, Anthony J and Farrelly, Terence C},
   year={2012},
   month=jan, pages={013063} 
}

@article{Mohapatra2025,
  title = {Exact Volume-Law Entangled Zero-Energy Eigenstates in a Large Class of Spin Models},
  author = {Mohapatra, Sashikanta and Moudgalya, Sanjay and Balram, Ajit C.},
  journal = {Phys. Rev. Lett.},
  volume = {134},
  issue = {21},
  pages = {210403},
  numpages = {7},
  year = {2025},
  month = {May},
  publisher = {American Physical Society},
  doi = {10.1103/PhysRevLett.134.210403},
  url = {https://link.aps.org/doi/10.1103/PhysRevLett.134.210403}
}

@misc{Perciavalle2025,
      title={Local reminiscence in the PXP model}, 
      author={Francesco Perciavalle and Gian Marco Rizzo and Francesco Plastina and Nicola Lo Gullo},
      year={2025},
      eprint={2509.19944},
      archivePrefix={arXiv},
      primaryClass={quant-ph},
      url={https://arxiv.org/abs/2509.19944}, 
}

@article{Rajagopal2010,
  title = {Kraus representation of quantum evolution and fidelity as manifestations of Markovian and non-Markovian forms},
  author = {Rajagopal, A. K. and Usha Devi, A. R. and Rendell, R. W.},
  journal = {Phys. Rev. A},
  volume = {82},
  issue = {4},
  pages = {042107},
  numpages = {7},
  year = {2010},
  month = {Oct},
  publisher = {American Physical Society},
  doi = {10.1103/PhysRevA.82.042107},
  url = {https://link.aps.org/doi/10.1103/PhysRevA.82.042107}
}

@article{Budini2025,
   title={Violation of diagonal non-invasiveness: A hallmark of non-classical memory effects},
   volume={555},
   ISSN={0375-9601},
   url={http://dx.doi.org/10.1016/j.physleta.2025.130788},
   DOI={10.1016/j.physleta.2025.130788},
   journal={Physics Letters A},
   publisher={Elsevier BV},
   author={Budini, Adrián A.},
   year={2025},
   month=sep, pages={130788} 
}

@ARTICLE{Ruskai1994,
       author = {{Ruskai}, Mary Beth},
        title = "{Beyond Strong Subadditivity? Improved Bounds on the Contraction of Generalized Relative Entropy}",
      journal = {Reviews in Mathematical Physics},
         year = 1994,
        month = jan,
       volume = {6},
        pages = {1147-1161},
          doi = {10.1142/S0129055X94000407}
}

@article{Chruscinski2011,
   title={Measures of non-Markovianity: Divisibility versus backflow of information},
   volume={83},
   ISSN={1094-1622},
   url={http://dx.doi.org/10.1103/PhysRevA.83.052128},
   DOI={10.1103/physreva.83.052128},
   number={5},
   journal={Physical Review A},
   publisher={American Physical Society (APS)},
   author={Chruściński, Dariusz and Kossakowski, Andrzej and Rivas, Angel},
   year={2011},
   month=may 
}

@article{Liu2013,
  title = {Nonunital non-Markovianity of quantum dynamics},
  author = {Liu, Jing and Lu, Xiao-Ming and Wang, Xiaoguang},
  journal = {Phys. Rev. A},
  volume = {87},
  issue = {4},
  pages = {042103},
  numpages = {6},
  year = {2013},
  month = {Apr},
  publisher = {American Physical Society},
  doi = {10.1103/PhysRevA.87.042103},
  url = {https://link.aps.org/doi/10.1103/PhysRevA.87.042103}
}

@ARTICLE{Jozsa1994,
       author = {{Jozsa}, Richard},
        title = "{Fidelity for Mixed Quantum States}",
      journal = {Journal of Modern Optics},
         year = 1994,
        month = dec,
       volume = {41},
       number = {12},
        pages = {2315-2323},
          doi = {10.1080/09500349414552171},
       adsurl = {https://ui.adsabs.harvard.edu/abs/1994JMOp...41.2315J}
}

\end{document}